\documentclass[english,american,aip,jcp,preprint]{revtex4}
\usepackage[T1]{fontenc}
\usepackage[latin9]{inputenc}
\usepackage{amsmath}
\usepackage{amssymb}
\usepackage{graphicx}
\usepackage{esint}

\makeatletter

\providecommand{\tabularnewline}{\\}

\@ifundefined{textcolor}{}
{%
 \definecolor{BLACK}{gray}{0}
 \definecolor{WHITE}{gray}{1}
 \definecolor{RED}{rgb}{1,0,0}
 \definecolor{GREEN}{rgb}{0,1,0}
 \definecolor{BLUE}{rgb}{0,0,1}
 \definecolor{CYAN}{cmyk}{1,0,0,0}
 \definecolor{MAGENTA}{cmyk}{0,1,0,0}
 \definecolor{YELLOW}{cmyk}{0,0,1,0}
}

\makeatother

\usepackage{babel}
\begin{document}

\title{{\normalsize Statistical mechanics of fluids confined by polytopes:
The hidden geometry of the cluster integrals}}

\author{{\normalsize Ignacio Urrutia\ref{CNEA-MC}}}

\affiliation{\label{CNEA-MC}Consejo Nacional de Investigaciones Científicas y
Técnicas, Argentina (CONICET) and Departamento de Física, Comisión
Nacional de Energía Atómica, Av. Gral. Paz 1499 (RA-1650) San Martín,
Buenos Aires, Argentina.}

\email{iurrutia@cnea.gov.ar}

\begin{abstract}
{\normalsize This paper, about a fluid-like system of spatially confined
particles, reveals the analytic structure for both, the canonical
and grand canonical partition functions. The studied system is inhomogeneously
distributed in a region whose boundary is made by planar faces without
any particular symmetry. This type of geometrical body in the $d$-dimensional
space is a polytope. The presented result in the case of $d=3$ gives
the conditions under which the partition function is a polynomial
in the volume, surface area, and edges length of the confinement vessel.
Equivalent results for the cases $d=1,2$ are also obtained. Expressions
for the coefficients of each monomial are explicitly given using the
cluster integral theory. Furthermore, the consequences of the polynomial
shape of the partition function on the thermodynamic properties of
the system, away from the so-called thermodynamic limit, is studied.
Some results are generalized to the $d$-dimensional case. The theoretical
tools utilized to analyze the structure of the partition functions
are largely based on integral geometry.}{\normalsize \par}
\end{abstract}

\date{{\normalsize \today}}

\maketitle

\section{{\normalsize Introduction\label{sec:Intro}}}

Thermodynamic properties of fluids are relevant to biology, chemistry,
physics and engineering. In most cases of interest the fluid system
is inhomogeneously distributed in the space, confined to a region
of finite size and constituted by a bounded number of molecules. A
typical example of this kind of systems is that of fluids confined
in pores. They constitute a prototypical inhomogeneous system which
occupies a small region of the space and may involve a small number
of particles. It is well known that the spatial distribution of a
fluid confined in a small pore may follow or not the symmetry of the
cavity \cite{Sartarelli_2010,Szybisz_2011}, and that its properties
may be strongly influenced by the geometry of the container. To study
the thermodynamic properties of these systems it is customary to introduce
several strong assumptions or approximations that simplify the analysis.
An usual approach is to treat the system as it were homogeneous (which
necessarily implies that the system completely fills the space, and
thus, involves infinitely many particles). A second usual approach
assumes that the inhomogeneous fluid is spatially distributed in several
regions, each one with homogeneous properties, while the inhomogeneous
nature of the system is concentrated in regions with vanishing size
(surfaces, lines and points). Frequently, this scenario is complemented
with the assumption that the spatial distribution of the fluid involves
continuous translational and/or rotational symmetries. Finally, a
third approach only assumes that the inhomogeneous fluid takes spatial
configurations with those symmetries. In general, the assumed symmetric
distribution of the fluid could be attained spontaneously (as in the
case of free drops or bubbles) or could be induced by an external
potential as it is the case of confined system that are constrained
to regions with simple symmetry: a semi-space, a slit, an infinite
cylinder, or a sphere (for example in wetting and capillary condensation,
phenomena). Anyway, the continuous symmetries (translations and/or
rotations) play an important role providing the bases to identify
the extensive and intensive magnitudes which enable the development
of thermodynamic theory \cite{Callen1985}.

Moreover, thermodynamics only provides an incomplete set of relations
between intensive and extensive magnitudes which must be complemented
with other sources of information to obtain the thermodynamic properties
of a given fluid system. In any case, these sources, that may be experimental,
theoretical, or based on numerical simulation, also involve assumptions
or approximations related to the existence of continuous symmetries.
In this sense, our statistical mechanical and thermodynamical approach
to the study of inhomogeneous fluids appears to be intrinsically entangled
with some hypothesis about the constitution and behavior of the system
under study %
\footnote{See Chapter 6 in Ref.\cite{Hansen2006} an also Ref.\cite{Henderson_1992_inHendersonD1992}.%
}. Particularly, I refer to three hypothesis concerning the fluid system:
the symmetry of its spatial distribution, the large volume occupied
and the large (and unbounded) number of particles involved. Of course,
one may ask about if those hypothesis are or not a central part of
the theories. What happens with inhomogeneous fluid-like systems which
do not necessary attain spatial distribution with simple symmetry
and/or occupy small regions of the space and/or are constituted by
a small number of particles? Can we apply statistical mechanics and
thermodynamics to study their equilibrium properties? How can we
do that? The analysis presented below attempts to advance in the
understanding of these questions. From a complementary point of view,
this work also deals with the long standing objective of finding fluid-like
systems that are exactly solvable, i.e., its partition function integral
can be integrated and thus transformed into an analytic expression.
Each of these systems provides the unique opportunity of testing some
of the fundamental hypothesis of the theoretical framework.

This paper is devoted to analyze, using an exact framework, few- and
many-body fluid like systems confined in cavities without continuous
symmetry. In particular, we focus on their partition function and
thermodynamic magnitudes as functions of the spatial set that defines
the region in which particles are allowed to move. In fact, it is
shown that under certain conditions the partition function of the
fluid-like system is a polynomial function of certain geometric measures
of the confining cavity, like its volume and surface area. The adopted
approach is theoretical and exact, being our core results largely
based on integral geometry. The rest of the manuscript is organized
as follows: In Section \ref{sec:PartitionFunc} it is presented the
general statistical mechanics approach to canonical and grand canonical
ensemble with two modifications: the cutoff in the maximum number
of particles which enables the analysis of few-body open systems and
the generalization of cluster integrals to inhomogeneous fluids. Sections
\ref{sec:The-properties-of-Taui} and \ref{sec:polytope-theo} are
devoted to study the structure of cluster integrals with particular
emphasis in the case of a polytope-type confinement. There, two theorems
and their corollaries are demonstrated, which constitute the main
result of the present work (PW). Extensions to other type of confinements
are discussed in Section \ref{sec:Extensions}, while the consequences
of the cluster integral behavior in the thermodynamical properties
of the system are studied in Sec. \ref{sec:Thermodynamics}. The final
discussion is given in Section \ref{sec:Conclu}.

\section{{\normalsize Partition function of an inhomogeneous system\label{sec:PartitionFunc}}}

In this work we consider an open system of at most $M$ particles
that evolves at constant temperature in a restricted region $\mathcal{A}$
of the $d$-dimensional euclidean space $\mathbb{R}^{d}$. The region
$\mathcal{A}$ is the set of points where the center of each particle
is free to move and has a boundary $\partial\mathcal{A}$. From here
on the system will be shortly referred to as the fluid while $\mathcal{A}$
will be referred to as the region or set where the fluid is confined.
In fact, the system is an inhomogeneous fluid due to the existence
of $\partial\mathcal{A}$. The restricted grand canonical partition
function of the fluid is
\begin{equation}
\Xi_{M}=\sum_{j=0}^{M}x^{j}Q_{j}\:,\label{eq:GPF}
\end{equation}
where $M$ is the maximum number of particles that are accepted in
$\mathcal{A}$. The absolute activity is $x=\exp\left(\beta\mu\right)$,
being $\mu$ the chemical potential while $\beta=1/k_{B}T$, $k_{B}$
and $T$ are the inverse temperature, Boltzmann constant and temperature,
respectively. $\Xi_{M}$ is a $M$-degree polynomial in $x$ and
its $j$-th coefficient is the canonical partition function of the
closed system with exactly $j$ particles confined in $\mathcal{A}$.
The grand canonical partition function of the fluid without the cutoff
in the maximum number of particles can be obtained from $\underset{M\rightarrow\infty}{\lim}\Xi_{M}$.
On the other hand, the canonical partition function of the system
with $j$ particles is
\begin{equation}
Q_{j}=\mathtt{I}_{j}\Lambda^{-d*j}Z_{j}\:,\label{eq:CPF}
\end{equation}
where $\mathtt{I}_{j}=1/j!$ is the indistinguishably factor, \foreignlanguage{english}{$\Lambda=h/(2\pi m\, k_{B}T)^{1/2}$}
is the thermal de Broglie wavelength, $m$ is the mass of each particle
and $h$ is the Planck's constant. Therefore, we can transform Eq.
(\ref{eq:GPF}) into
\begin{equation}
\Xi_{M}=\sum_{j=0}^{M}\mathtt{I}_{j}z^{j}Z_{j}\:,\label{eq:GPF2}
\end{equation}
where the activity is $z=\exp\left(\beta\mu\right)/\Lambda^{d}$.
Finally, the configuration integral (CI) of a system with $j$ particles
is
\begin{equation}
Z_{j}=\int\ldots\int\,\prod_{k=1}^{j}e_{k}\prod_{<l,n>}e_{ln}\, d^{j}\mathbf{r}\:,\label{eq:Zjinteg}
\end{equation}
being $Z_{0}=1$ {[}alternative dimensionless definitions for $z$
and $Z_{j}$ may be obtained by introducing the volume of the system
or the characteristic volume of a particle in Eq. (\ref{eq:Zjinteg}){]}.
Here, $e_{ln}=\exp\left[-\beta\phi(r_{ln})\right]$ is the Boltzmann
factor related to the spherically symmetric pair potential $\phi$
between the $l$ and $n$ particles, which are separated by a distance
$r_{ln}$. It will be assumed that $\phi$ has a finite range $\xi$,
being $\phi(r_{ln})=0$ and $e_{ln}=1$ if $r_{ln}>\xi$. This assumption
is not very restrictive because any pair interaction potential can
be approximated by truncation at a finite range, e.g. it is frequent
to study the Lennard Jones fluid cut and shifted at $r=2.5\sigma$
\cite{Shaul_2010,Horsch_2012}. The indicator function
\begin{eqnarray}
e_{k}=e(\mathbf{r}_{k}) & = & \begin{cases}
1 & \textrm{if }\mathbf{r}_{k}\in\mathcal{A}\:,\\
0 & \textrm{if }\mathbf{r}_{k}\notin\mathcal{A}\:,
\end{cases}\label{eq:el0}
\end{eqnarray}
is the Boltzmann factor corresponding to the external potential produced
by a hard wall confinement. Note that, the integration domain in Eq.
(\ref{eq:Zjinteg}) is the complete space due to the spatial confinement
of the particles in $\mathcal{A}$ is considered through $e_{k}$.
The CI is a function of the set $\mathcal{A}$ and a functional of
$\phi$. $Z_{i}$ itself is given in its full generality by \cite{Hill1956}
\begin{equation}
Z_{j}=j!\sum_{\mathbf{m}}\left[\prod_{i=1}^{j}\frac{1}{m_{i}!}\left(\frac{\tau_{i}}{i!}\right)^{m_{i}}\right]\:,\label{eq:ZjTaui}
\end{equation}
where the sum is over all sets of positive integers or zero $\mathbf{m}=\{m_{1},m_{2},\ldots\}$
such that $\sum_{i=1}^{j}i\, m_{i}=j$. Eq. (\ref{eq:ZjTaui}) shows
that $Z_{j}$ is a polynomial in $\tau_{1},\ldots,\tau_{j}$ which
are essentially the (reducible) Mayer cluster integrals for inhomogeneous
systems. This equation is obtained through the following procedure:
replace in Eq. (\ref{eq:Zjinteg}) each $e_{ln}$ using the identity
$e_{ln}=1+f_{ln}$ (this Eq. defines $f_{ln}$), distribute the products,
and collect all terms of the integrand which consist of groups of
particles that conform a cluster, in the sense that they are at least
simply connected between them by $f$-functions. Note that our assumption
about $e_{ln}=1$ for $r_{ln}>\xi$ implies $f_{ln}=0$ if $r_{ln}>\xi$
showing the spatial meaning of the cluster term. The procedure gives
Eq. (\ref{eq:ZjTaui}) with
\begin{equation}
\tau_{i}=\dotsint\, S_{1,2,\ldots,,i}\prod_{k=1}^{i}e_{k}d^{i}\mathbf{r}\:,\label{eq:Taui}
\end{equation}
\begin{equation}
S_{1,2,\ldots,i}=\sum_{cluster}\prod_{<l,n>}f_{ln}\:,\label{eq:S1i}
\end{equation}
where $S_{1,2,\ldots,i}$ is a sum of products of $f_{ln}$ functions
with $1\leq l,n\leq i$ that involves all the products of $f$ functions,
which can be represented as a connected diagrams (clusters) with $i$
nodes and $f_{ln}$ bonds. Clearly, $\tau_{i}$ depends on $T$, $\phi$
and $\mathcal{A}$. If one assume that $\mathcal{A}$ is very large
the usual homogeneous system approximation gives $\tau_{i}\rightarrow i!Vb_{i}$,
where $V$ is the volume of $\mathcal{A}$ and $b_{i}$s are the Mayer
cluster integrals which depend on $T$ and $\phi$.

The inversion of Eq. (\ref{eq:ZjTaui}) gives the following expression
for the dependence of $\tau_{i}$ with the CIs (Eq. (23.44) in Ref.
\cite{Hill1956})
\begin{equation}
\tau_{i}=i!\sum_{\mathbf{n}}\left(-1\right)^{\sum_{j}n_{j}-1}\left(\sum_{j}n_{j}-1\right)!\left[\prod_{j=1}\frac{1}{n_{j}!}\left(\frac{Z_{j}}{j!}\right)^{n_{j}}\right]\:,\label{eq:TauiZj}
\end{equation}
where the first sum is over all sets $\mathbf{n}=\left\{ n_{1},n_{2},\ldots\right\} $
with $n_{j}$ non-negative integers such that $\sum_{j=1}^{i}j\, n_{j}=i$
and $\tau_{1}=Z_{1}b_{1}$ with $b_{1}=1$. From Eq. (\ref{eq:TauiZj})
it is apparent that $\tau_{i}$ depends on $Z_{j}$ with $j=1,\ldots,i$,
and thus, the $\tau_{i}$s with $i=1,...,k$ and the $Z_{j}$s with
$j=1,\ldots,k$ involve the same physical information. On the other
hand, one can return to the Eqs. (\ref{eq:GPF}) - (\ref{eq:Zjinteg})
to observe that they can be re-written in this alternative form: replace
$M\rightarrow\infty$ in Eq. (\ref{eq:GPF}) but assume $Z_{M+k}=0$
with $k=1,\ldots,\infty$. In this context Eq. (\ref{eq:TauiZj})
shows that cluster integrals $\tau_{i}$ with $i=1,...,M$ are not
affected by the restriction $j\leq M$, being $\tau_{M+1}$ the first
affected $\tau$ (because it depends on $Z_{M+1}$, which is zero).

Before ending this section we wish to focus on a relevant characteristic
of $\Xi_{M}$, $Z_{j}$ and $\tau_{i}$ functions. Let us define $\mathbb{S}=\left\{ \mathcal{A}/\mathcal{A}\subseteq\mathbb{R}^{d}\right\} $
(the set of all the subsets of $\mathbb{R}^{d}$), for fixed $T$,
$\phi$ and $z$ one can write $\Xi_{M}(\mathcal{A}):\mathbb{S}\rightarrow\mathbb{R}$
which implies that $\Xi_{M}(\mathcal{A})$ may depend on the shape
of $\mathcal{A}$. Clearly, the same argument applies to $Z_{j}$
and $\tau_{i}$ which may also depend on the shape of $\mathcal{A}$.
Here we anticipate the principal result of PW, related with this non-trivial
shape dependence, that will be demonstrated in the Secs. \ref{sec:The-properties-of-Taui}
and \ref{sec:polytope-theo}. Thus, we turn the attention to a fluid
confined by a polytope $\mathcal{A}$. For simplicity we focus in
the three dimensional case, i.e. a fluid confined by a polyhedron.
If a system of particles that interact via a pair potential of finite
range $\xi$ is confined in a polyhedron $\mathcal{A}$ such that
its characteristic length $\mathfrak{L}(\mathcal{A})$ {[}see Eq.
(\ref{eq:LdeA}){]} is greater than  \foreignlanguage{english}{$k\xi+C$}
(being C a constant) for some integer $k\geq1$. Then the $i$-th
cluster integral $\tau_{i}$ with $1\leq i\leq k$ is a linear function
in the variables $V$, $A$ and $\{L_{1},L_{2},...\}$ (the length
of the edges of $\mathcal{A}$). In fact
\begin{equation}
\tau_{i}/i!=V\, b_{i}-\, A\, a_{i}+\sum_{n\textrm{edges}}L_{n}c_{i,n}^{\textrm{e}}+\sum_{n\textrm{vertex}}c_{i,n}^{\textrm{v}}\;,\label{eq:TauiCorol-pre}
\end{equation}
where the coefficients $b_{i}$ and $a_{i}$ are independent of the
shape of $\mathcal{A}$ while $c_{i,n}^{\textrm{e}}$ and $c_{i,n}^{\textrm{v}}$
are functions of the dihedral angles involved. Besides, all the coefficients
depend on the pair interaction potential and temperature. Expressions
similar to Eq. (\ref{eq:TauiCorol-pre}) are also found for the euclidean
space with dimension $2$ and $1$, while they are conjectured for
dimension larger than $3$. Two non-trivial consequences derive from
the Eq. (\ref{eq:TauiCorol-pre}). On one hand it implies that, if
the system of particles confined by $\mathcal{A}$ involves $N$ particles
and $\mathfrak{L}(\mathcal{A})>N\xi+C$, then $Z_{N}$ is polynomial
on $V$, $A$ and $\{L_{1},L_{2},...\}$. On the other hand it implies
that, if the system of particles confined by $\mathcal{A}$ is open,
involves at most $M$ particles and $\mathfrak{L}(\mathcal{A})>M\xi+C$,
then $\Xi_{M}$ is a polynomial in $z$, $V$, $A$ and $\{L_{1},L_{2},...\}$.

\section{{\normalsize The properties of some functions related to $\tau_{i}$\label{sec:The-properties-of-Taui}}}

In this section we analyze the properties of some many-body functions
related to the partial integration of the cluster integral $\tau_{i}$
{[}Eqs. (\ref{eq:Taui}) and (\ref{eq:S1i}){]}. This analysis will
be complemented in the next section where we will reveal the linear
behavior of $\tau_{i}$. In the following paragraphs several definitions
are introduced and two different proofs of the locality and rigid
invariance of functions related to the partial integration of $\tau_{i}$
are presented. Both proofs are necessary for clarity. In the first
approach the mentioned properties of those functions are demonstrated
and it is found a cutoff for its finite range, while in the second
approach (which is more complex than the first one) the properties
are proved and a better bound of the finite range is obtained.

We define, $G(\mathcal{A},\mathbf{r}):\mathbb{S}\times\mathbb{R}^{d}\rightarrow\mathbb{R}$
is a local function in $\mathbf{r}$ over $\mathcal{A}$ with range
$\lambda$ (from hereon local with range $\lambda$) if its value
is entirely determined by the set $U(\mathbf{r},\lambda)\cap\mathcal{A}$
that is $G(\mathcal{A},\mathbf{r})=G(U(\mathbf{r},\lambda)\cap\mathcal{A})$
where $U(\mathbf{r},\lambda)$ is the ball centered at $\mathbf{r}$
with radius $\lambda$. This definition can be generalized to functions
of several variables in the following way: $G(\mathcal{A},\mathbf{r}_{1},\ldots,\mathbf{r}_{n}):\mathbb{S}\times\mathbb{R}^{d*n}\rightarrow\mathbb{R}$
is said to be a local function in $\mathbf{r}_{i}$ over $\mathcal{A}$
with range $\lambda$ if $G(\mathcal{A},\mathbf{r}_{1},\ldots,\mathbf{r}_{n})=G(U(\mathbf{r}_{i},\lambda)\cap\mathcal{A},\mathbf{r}_{1},\ldots,\mathbf{r}_{i-1},\mathbf{r}_{i+1},\ldots,\mathbf{r}_{n})$
for $\left\{ \mathbf{r}_{1},\ldots,\mathbf{r}_{i-1},\mathbf{r}_{i+1},\ldots,\mathbf{r}_{n}\right\} \subset U(\mathbf{r}_{i},\lambda)\cap\mathcal{A}$
and $G(\mathcal{A},\mathbf{r}_{1},\ldots,\mathbf{r}_{n})=0$ when
exist $\mathbf{r}_{j\neq i}\notin U(\mathbf{r}_{i},\lambda)\cap\mathcal{A}$.

Let $G(\mathcal{A},\mathbf{r})$ be a local function with range $\lambda$.
We say that $G(\mathcal{A},\mathbf{r})$ is invariant under rigid
transformations (the elements of the euclidean group, i.e., any composition
of translations, rotations, inversions and reflections) if $\forall$
rigid transformation $R$, $R[U(\mathbf{r},\lambda)\cap\mathcal{A}]=U(\mathbf{r}',\lambda)\cap\mathcal{A}'$
with $\mathbf{r}'=R\mathbf{r}$ implies $G[U(\mathbf{r},\lambda)\cap\mathcal{A}]=G[U(\mathbf{r}',\lambda)\cap\mathcal{A}']$.
The generalization to functions with many variables is: let $G(\mathcal{A},\mathbf{r}_{1},\ldots,\mathbf{r}_{n}):\mathbb{S}\times\mathbb{R}^{d*n}\rightarrow\mathbb{R}$
be a local function in $\mathbf{r}_{i}$ with range $\lambda$, we
say that $G(\mathcal{A},\mathbf{r}_{1},\ldots,\mathbf{r}_{n})$ is
invariant under rigid transformations if $\forall$ rigid transformation
$R$, $R[U(\mathbf{r}_{i},\lambda)\cap\mathcal{A}]=U(\mathbf{r}_{i}',\lambda)\cap\mathcal{A}'$
implies $G[U(\mathbf{r}_{i},\lambda)\cap\mathcal{A},\mathbf{r}_{1},\ldots]=G[U(\mathbf{r}_{i}',\lambda)\cap\mathcal{A}',\mathbf{r}_{1}',\ldots]$,
with $\mathbf{r}_{j}'=R\mathbf{r}_{j}$. A direct consequence of the
locality with range $\lambda$ and rigid transformation invariance
of a bounded function $G(\mathcal{A},\mathbf{r})$ is that it attains
a constant value for all $\mathbf{r}\in\mathcal{A}$ such that $U(\mathbf{r},\lambda)\cap\partial\mathcal{A}$.

\medskip{}

\label{Th1}Theorem 1: Let $e_{k}(\mathbf{r})$ and $S_{1,2,\ldots,i}$
be the Boltzmann factor and cluster integrand introduced in Eqs. (\ref{eq:el0})
to (\ref{eq:S1i}), and $\mathcal{A}\in\mathbb{S}$ a set in $\mathbb{R}^{d}$
for which the integral in Eq. (\ref{eq:E1}) is finite. Then
\begin{equation}
E_{1}(\mathcal{A},\mathbf{r}_{1})\equiv\dotsint\, S_{1,2,\ldots,i}\prod_{k=2}^{i}e_{k}d\mathbf{r}_{2}\cdots d\mathbf{r}_{i}\:\label{eq:E1}
\end{equation}
is a local function with finite range invariant under rigid transformations. 

\medskip{}

\label{Aproof}First Proof: Consider the following $i$-body function
\begin{equation}
E_{i}(\mathbf{r}_{1},\mathbf{y}_{2},\ldots,\mathbf{y}_{i})=S_{1,2,\ldots,i}\:,\label{eq:EconS}
\end{equation}
with $\mathbf{r}_{1}$ the position of an arbitrarily chosen particle,
and the rule to obtain the $(n-1)$-body function from the $n$-body
one given by 
\begin{equation}
E_{n-1}(\mathcal{A},\mathbf{r}_{1},\mathbf{y}_{2},\ldots,\mathbf{y}_{n-1})\equiv\int\, E_{n}(\mathcal{A},\mathbf{r}_{1},\mathbf{y}_{2},\ldots,\mathbf{y}_{n})\, e_{n}d\mathbf{y}_{n}\:,\label{eq:E2fun}
\end{equation}
where $2\leq n\leq i$ and $\mathbf{y}_{j}=\mathbf{r}_{j}-\mathbf{r}_{1}$
is the coordinate of particle $j$ with respect to particle $1$.
$S_{1,2,\ldots,i}$ has range
\begin{equation}
\varsigma=\left(i-1\right)\xi\:,\label{eq:psi1}
\end{equation}
being $S_{1,2,\ldots,i}=0$ if $r_{ab}=\left|\mathbf{r}_{a}-\mathbf{r}_{b}\right|>\left(i-1\right)\xi$
for at least one pair of particles $a$ and $b$ in the cluster. This
property derives from the fact that $S_{1,2,\ldots,i}$ contains the
open simple-chain cluster term $S'=f_{12}f_{23}\cdots f_{i-1,i}$
that enables that two particles reach the maximum possible separation
$\left(i-1\right)\xi$ of all the cluster terms in $S_{1,2,\ldots,i}$
(there are $i!/2$ terms of this type). Even more, one can show that
$S_{1,2,\ldots,i}$ is local with range $\varsigma$. Naturally, Eq.
(\ref{eq:EconS}) shows that $E_{i}(\mathbf{r}_{1},\mathbf{y}_{2},\ldots,\mathbf{y}_{i})$
is also local with range $\varsigma$. The integration in Eq. (\ref{eq:E2fun})
applied to $E_{i}(\mathbf{r}_{1},\mathbf{y}_{2},\ldots,\mathbf{y}_{i})$
implies that $E_{i-1}(\mathcal{A},\mathbf{r}_{1},\mathbf{y}_{2},\ldots,\mathbf{y}_{i-1})$
is local in $\mathbf{r}_{1}$ with range $\varsigma$. We proceed
by induction. Let us assume that for some $n<i$, $E_{n}(\mathcal{A},\mathbf{r}_{1},\mathbf{y}_{2},\ldots,\mathbf{y}_{n})$
is local with range $\varsigma$, then by Eq. (\ref{eq:E2fun}) $E_{n-1}(\mathcal{A},\mathbf{r}_{1},\mathbf{y}_{2},\ldots,\mathbf{y}_{n-1})$
is also local with the same range. The procedure continue until the
function $E_{1}(\mathcal{A},\mathbf{r}_{1})$ is reached. Therefore,
we find that the $E_{n}(\mathcal{A},\mathbf{r}_{1},\mathbf{y}_{2},\ldots,\mathbf{y}_{n})$
functions with $1\leq n<i$ are local in \foreignlanguage{english}{$\mathbf{r}_{1}$}
with range $\varsigma$ {[}even more, we have obtained that $E_{n}(\mathcal{A},\mathbf{r}_{1},\ldots,\mathbf{r}_{n})$
are local in \foreignlanguage{english}{$\mathbf{r}_{j}$} with range
$\varsigma$, for any $1\leq j\leq n${]}.

Taking into account that $E_{i}(\mathbf{r}_{1},\mathbf{y}_{2},\ldots,\mathbf{y}_{i})$
has range $\varsigma$ and that $e_{i}=e_{i}(\mathcal{A},\mathbf{r}_{i})$
{[}see Eq. (\ref{eq:el0}){]} the right hand side of Eq. (\ref{eq:E2fun})
for $n=i$ can be written as
\begin{equation}
\int_{U(\mathbf{r}_{1},\varsigma)\cap\mathcal{A}}\, E_{i}(\mathbf{r}_{1},\mathbf{y}_{2},\ldots,\mathbf{y}_{i})\, d\mathbf{y}_{i}\:.\label{eq:iurt01}
\end{equation}
It is convenient to express the coordinates $(\mathbf{r}_{1},\mathbf{y}_{2},\ldots,\mathbf{y}_{i})$
in terms of the rigid transformed coordinates $(\mathbf{r}_{1}',\mathbf{y}_{2}',\ldots,\mathbf{y}_{i}')$
(related each other by $\mathbf{r}_{j}=R^{-1}\mathbf{r}_{j}'$ and
$\mathbf{y}_{j}=R^{-1}\mathbf{y}_{j}'=R^{-1}\mathbf{r}_{j}'-R^{-1}\mathbf{r}_{1}'$
with $R^{-1}$ the inverse of $R$) and to change the integration
variable to $\mathbf{y}_{i}'$ (note that the Jacobian is one). Thus
we find 
\begin{equation}
\int_{R\left[U(\mathbf{r}_{1},\varsigma)\cap\mathcal{A}\right]}\, E_{i}(R^{-1}\mathbf{r}_{1}',R^{-1}\mathbf{y}_{2}',\ldots,R^{-1}\mathbf{y}_{i}')\, d\mathbf{y}_{i}'\:,\label{eq:iurt02}
\end{equation}
where the integration domain is equal to $U(\mathbf{r}_{1}',\varsigma)\cap\mathcal{A}'$
by hypothesis. Given that $E_{i}$ is invariant under any rigid transformation
applied to the coordinate of the particles one can drop each $R^{-1}$
in Eq. (\ref{eq:iurt02}) and return to the original form introducing
$e_{i}=e_{i}(\mathcal{A}',\mathbf{r}_{i}')$ 
\begin{equation}
\int\, E_{i}(\mathbf{r}_{1}',\mathbf{y}_{2}',\ldots,\mathbf{y}_{i}')\, e_{i}\, d\mathbf{y}_{i}'\:,\label{eq:iurt03}
\end{equation}
which is the definition of $E_{i-1}(\mathcal{A}',\mathbf{r}_{1}',\mathbf{y}_{2}',\ldots,\mathbf{y}_{i-1}')$.
This shows that $E_{i-1}(\mathcal{A},\mathbf{r}_{1},\mathbf{y}_{2},\ldots,\mathbf{y}_{i-1})$
is invariant under rigid transformations. Again, we proceed by induction.
Let us assume that $E_{n}(\mathcal{A},\mathbf{r}_{1},\mathbf{y}_{2},\ldots,\mathbf{y}_{n})$,
a local function in $\mathbf{r}_{1}$ over $\mathcal{A}$ with range
$\varsigma$, is invariant under rigid transformations. Taking into
account that $e_{n}=e_{n}(\mathcal{A},\mathbf{r}_{n})$ for the right
hand side of Eq. (\ref{eq:E2fun}) we can write an expression similar
to Eq. (\ref{eq:iurt01}) but replacing $E_{i}(\mathbf{r}_{1},\mathbf{y}_{2},\ldots,\mathbf{y}_{i})$
by $E_{n}(U(\mathbf{r}_{1},\varsigma)\cap\mathcal{A},\mathbf{y}_{2},\ldots,\mathbf{y}_{n})$
and $i$ by $n$. Turning to transformed coordinates we find 
\begin{equation}
\int_{R\left[U(\mathbf{r}_{1},\varsigma)\cap\mathcal{A}\right]}\, E_{n}(R^{-1}[U(\mathbf{r}_{1}',\varsigma)\cap\mathcal{A}'],R^{-1}\mathbf{y}_{2}',\ldots,R^{-1}\mathbf{y}_{n}')\, d\mathbf{y}_{n}'\:,\label{eq:iurt04}
\end{equation}
where the integration domain is equal to $U(\mathbf{r}_{1}',\varsigma)\cap\mathcal{A}'$
and we used that $U(\mathbf{r}_{1},\varsigma)\cap\mathcal{A}=R^{-1}[U(\mathbf{r}_{1}',\varsigma)\cap\mathcal{A}']$.
Given that $E_{n}$ is invariant under rigid transformations we can
drop each $R^{-1}$ in the arguments of $E_{n}$ in Eq. (\ref{eq:iurt04}),
use the finite range of $E_{n}$ to split the argument $U(\mathbf{r}_{1}',\varsigma)\cap\mathcal{A}'$
into $(\mathcal{A}',\mathbf{r}_{1}')$ and introduce $e_{n}(\mathcal{A}',\mathbf{r}_{n}')$
to obtain
\begin{equation}
\int\, E_{n}(\mathcal{A}',\mathbf{r}_{1}',\mathbf{y}_{2}',\ldots,\mathbf{y}_{n}')\, e_{n}\, d\mathbf{y}_{n}'\:,\label{eq:iurt05}
\end{equation}
which is the definition of $E_{n-1}(\mathcal{A}',\mathbf{r}_{1}',\mathbf{y}_{2}',\ldots,\mathbf{y}_{n-1}')$.
Therefore, $E_{n-1}\left(\mathcal{A},\mathbf{r}{}_{1},\mathbf{y}{}_{2},\ldots,\mathbf{y}{}_{n-1}\right)$
is local of range $\varsigma$ and rigid transformation invariant.
The procedure continue until the function $E_{1}$ is reached. Therefore,
returning to the original coordinates, we obtain that $E_{n}\left(\mathcal{A},\mathbf{r}{}_{1},\ldots,\mathbf{r}{}_{n}\right)$
with $1\leq n\leq i$ is local in any $\mathbf{r}{}_{j}$ ($1\leq j\leq n$)
with range $\varsigma$ and rigid transformation invariant. In particular,
for $n=1$ it implies that $E_{1}(\mathcal{A},\mathbf{r}_{1})$ is
local in \foreignlanguage{english}{$\mathbf{r}_{1}$} with range $\varsigma$
and rigid transformation invariant.\hfill{}$\blacksquare$

\medskip{}

\label{Bproof}Second Proof: This second proof is based on a more
subtle consideration about the particle labeled as $\mathbf{r}{}_{1}$
in the first proof. We introduce the coordinate of the central particle
of the cluster, $\mathbf{r}_{\mathbf{c}}$, and the relative coordinates
$\mathbf{y}_{j}=\mathbf{r}_{j}-\mathbf{r}_{\textrm{c}}$ of the particle
$j$ with respect to $\mathbf{r}_{\textrm{c}}$. Again, consider the
cluster integral $\tau_{i}$ and its simple open-chain cluster $S'$
term. For $S'$ we take $\mathbf{r}_{\textrm{c}}=\mathbf{r}_{l}$
with $l=\textrm{IntegerPart}[i/2+1]$ (if $i$ is an odd number $\mathbf{r}_{\textrm{c}}=\mathbf{r}_{(i+1)/2}$
is the position of the middle-chain particle while if $i$ is an even
number $\mathbf{r}_{\textrm{c}}=\mathbf{r}_{i/2+1}$ is the position
of one of the pair of particles that are at the middle of the chain).
We note that $S'_{1,2,\ldots,i}$ is zero if $\left|\mathbf{r}_{a}-\mathbf{r}_{\textrm{c}}\right|>\left(l-1\right)\xi$
for at least one particle $a$ in the cluster. The procedure to obtain
$\mathbf{r}_{\textrm{c}}$ for all the other terms in $S_{1,2,\ldots,i}$
is as follows: for a given cluster $S''$ of $i$ particles take iteratively
each pair of particles, separate them to find the maximum possible
elongation distance under the condition $S''\neq0$. Let the more
stretchable chain of particles (that could be non-unique) be a chain
of $k$ particles with end-particles $a$ and $b$. Thus, $\left|\mathbf{r}_{a}-\mathbf{r}_{b}\right|<\left(k-1\right)\xi$
with $2\leq k<i$ and for this cluster we can define $\mathbf{r}_{\textrm{c}}=\mathbf{r}_{l}$
with $l=\textrm{IntegerPart}[k/2+1]$. By using this second approach
we find that $S_{1,2,\ldots,i}$ has finite range
\begin{equation}
\varsigma=\left(i-1\right)\xi/2\:\textrm{if }i\textrm{ is odd},\:\:\varsigma=i\xi/2\:\textrm{if }i\textrm{ is even}\:.\label{eq:psi2}
\end{equation}
Once $\mathbf{r}_{\textrm{c}}$ is identified for each cluster term
in $S_{1,2,\ldots,i}$ we can rename $\mathbf{r}_{\textrm{c}}$ as
$\mathbf{r}_{1}$ and follow the procedure developed in the first
proof of the theorem.\hfill{}$\blacksquare$

Based on Eq. (\ref{eq:psi2}) along with the assumption that the range
of $E(\mathbf{r})$ must be a unique function of the maximum elongation
length of $S_{1,2,\ldots,i}$ independently of the parity of $i$
we have the following guess
\begin{equation}
\varsigma=\left(i-1\right)\xi/2\;\;\:\forall i\in\mathbb{N}\:.\label{eq:psi3guess}
\end{equation}
We may mention that Eq. (\ref{eq:psi3guess}) can be demonstrated
for the case of $\mathcal{A}$ being a convex body {[}by virtue of
Eq. (\ref{eq:psi2}) one must focus on the case of even $i${]}. The
demonstration follows a procedure similar to that used to obtain the
Eq. (\ref{eq:psi2}), but for the case of even $i$ one define $\mathbf{r}_{\textrm{c}}=\left(\mathbf{r}_{i/2}+\mathbf{r}_{(i/2+1)}\right)/2$.
The development made in this section enable to write Eq. (\ref{eq:Taui})
as
\begin{equation}
\tau_{i}=\int E(\mathbf{r}_{1})e_{1}d\mathbf{r}_{1}\:,\label{eq:TauiE}
\end{equation}
with two different definitions for $E(\mathbf{r}_{1})$. Both definitions
and any other possible approach must give mathematically equivalent
expressions for $\tau_{i}$.

\section{{\normalsize Integration over a polytope-shaped domain\label{sec:polytope-theo}}}

In order to analyze the implications of the domain's shape on certain
type of integrals it is necessary to introduce some notions about
sets $\mathcal{B},\mathcal{C}\subseteq\mathbb{R}^{d}$. They are summarized
in this and the next paragraphs. The closure of $\mathcal{B}$ is
$\textrm{cl}\left(\mathcal{B}\right)=\mathcal{B}\cup\partial\mathcal{B}$.
It is said that $\mathcal{B}$ is a closed set if $\mathcal{B}=\textrm{cl}\left(\mathcal{B}\right)$.
Besides, the interior of $\mathcal{B}$ is $\textrm{int}\left(\mathcal{B}\right)=\mathcal{B}\setminus\partial\mathcal{B}$.
It is said that $\mathcal{B}$ is an open set if $\mathcal{B}=\textrm{int}\left(\mathcal{B}\right)$.
There are sets that are neither closed nor open. $\left\{ \mathcal{B}_{1},\ldots,\mathcal{B}_{n}\right\} $
is a partition of the non-empty set $\mathcal{B}$ if $\mathcal{B}=\bigcup_{k=1}^{n}\mathcal{B}_{k}$
and, for all $i,j\in\mathbb{N}_{\leq n}$ with $i\neq j$, $\mathcal{B}_{i}\neq\emptyset$
and $\mathcal{B}_{i}\cap\mathcal{B}_{j}=\emptyset$. A set in $\mathbb{R}^{d}$
is connected if it cannot be partitioned in two non-empty sets $\mathcal{B}$
and $\mathcal{C}$ such that $\textrm{cl}\left(\mathcal{B}\right)\cap\mathcal{C}=\mathcal{B}\cap\textrm{cl}\left(\mathcal{C}\right)=\emptyset$,
otherwise it is disconnected. We also introduce the concept of connectedness-based
partition of a set; a partition $\left\{ \mathcal{B}_{1},\ldots,\mathcal{B}_{n}\right\} $
of $\mathcal{B}$ is the connectedness-based partition of $\mathcal{B}$
if, for all $i,j\in\mathbb{N}_{\leq n}$ with $i\neq j$, $\mathcal{B}_{i}$
is connected and $\mathcal{B}_{i}\cup\mathcal{B}_{j}$ is disconnected.
The following notions of distances are adopted: the distance between
the point $\mathbf{r}\in\mathbb{R}^{d}$ and the set $\mathcal{B}$
is $\textrm{d}\left(\mathbf{r},\mathcal{B}\right)=\min\left(\left|\mathbf{r}-\mathbf{b}\right|,\mathbf{b}\in\mathcal{B}\right)$
with $\left|\mathbf{r}-\mathbf{b}\right|$ the usual Euclidean distance
between points, while the distance between the sets $\mathcal{B}$
and $\mathcal{C}$ is $\textrm{d}\left(\mathcal{B},\mathcal{C}\right)=\min\left(\left|\mathbf{b}-\mathbf{c}\right|,\mathbf{b}\in\mathcal{B}\textrm{ and }\mathbf{c}\in\mathcal{C}\right)$.
Finally, a set $\mathcal{B}$ is said to be convex if every pair of
points $x$ and $y$ in $\mathcal{B}$ are the endpoints of a line
segment lying inside $\mathcal{B}$.

An affine $k$-subspace is a linear variety of rank $k$ in $\mathbb{R}^{d}$,
whether it contains the origin or not. For example, the $\left(d-1\right)$-dimensional
affine subspace is a hyperplane. The affine hull of a set $\mathcal{B}\subseteq\mathbb{R}^{d}$,
$\textrm{aff}\left(\mathcal{B}\right)$, is the affine subspace with
the smallest rank in which every point of $\mathcal{B}$ is contained.
Given that $d$ is a fixed parameter from hereon we will refer to
the rank of $\textrm{aff}\left(\mathcal{B}\right)$ as the dimension
of the affine space of $\mathcal{B}$, denoted $\textrm{dim}\left[\textrm{aff}\left(\mathcal{B}\right)\right]$
or simply $\textrm{dim}\left(\mathcal{B}\right)$. For $\mathcal{B}\subseteq\mathbb{R}^{d}$,
the sets $\textrm{relcl}\left(\mathcal{B}\right)$, $\textrm{relint}\left(\mathcal{B}\right)$
and $\textrm{rel\ensuremath{\partial}}\left(\mathcal{B}\right)$ are
the relative closure, the relative interior and the relative boundary
of $\mathcal{B}$, i.e., the closure, interior and boundary of $\mathcal{B}$
within its affine hull, respectively \cite{Schneider2008}.

Even though a polytope is essentially a mathematical entity, in the
current work it also has a physical meaning because it serves to characterize
the shape of the vessel where particles are confined. This duality
makes difficult to start with a simple and satisfactory geometrical
description of the kind of polytopes that are relevant for our purposes.
We begin a somewhat indirect approach that ends in the definition
of the set of polytopes that have simple boundary (a well behaved
one), in the sense that the boundary can be dissected in its elements
or faces. To make further progress it is convenient to introduce the
convex polytopes which, from a geometrical point of view, are simpler
than the general polytopes. A convex polytope in $\mathbb{R}^{d}$
is any set with non-null volume given by the intersection of finitely
many half-spaces \cite{Schneider2008} (this definition includes unbounded,
closed and non-closed polytopes). On the other hand, a connected and
closed (CC) polytope $\mathcal{A}$ is the connected union of finitely
many closed convex polytopes. We define, the CC polytope $\mathcal{A}\subseteq\mathbb{R}^{d}$
is a simple-boundary (SB) polytope if for each $\mathbf{r}\in\partial\mathcal{A}$
$\exists$ $\varepsilon\in\mathbb{R}_{>0}$ such that $\forall\lambda\,0<\lambda<\varepsilon$,
both sets $\textrm{int}\left[U\left(\mathbf{r},\lambda\right)\cap\mathcal{A}\right]$
and $\textrm{int}\left[U\left(\mathbf{r},\lambda\right)\cap\mathcal{A}^{c}\right]$
are topologically equivalent (homeomorphic) to an open ball. It
is clear that a closed convex polytope is also a CC SB polytope. In
this sense, we say that the boundary of a CC SB polytope and the boundary
of a convex polytope are locally equivalent. The SB condition excludes
some degenerate or pathological cases e.g. a polytope with two vertex
or two edges, in contact. On the other hand, the definition of SB
polytopes includes bounded and unbounded polytopes, non-convex polytopes,
polytopes with holes or cavities and many kind of faceted knots embedded
in $\mathbb{R}^{d}$. Let $\mathbb{P}^{d}$ be the class of CC SB
polytopes in $\mathbb{R}^{d}$.

For a given polytope we focus on the partition of its boundary based
on its faces. As before, we treat first the convex polytope case.
Let $\mathcal{B}\in\mathbb{P}^{d}$ be a convex polytope and $H$
a hyperplane. If $H\cap\partial\mathcal{B}\neq\emptyset$, $H\cap\textrm{int}\left(\mathcal{B}\right)=\emptyset$
and $k=\textrm{dim}\left(H\cap\partial\mathcal{B}\right)$ with $0\leq k\leq d-1$,
then we say that $H\cap\partial\mathcal{B}$ is a closed $k$-face
of $\mathcal{B}$ \cite{Grunbaum_1969}. One can demonstrate that
the set whose elements are all the closed $0$-faces and relopen $k$-faces
with $0<k\leq d-1$, of $\mathcal{B}$ is a partition of $\partial\mathcal{B}$.
Other polytopes, non-necessarily convex, may admit a similar face-decomposition
of its boundary. Given a polytope $\mathcal{A}$ with face-decomposable
boundary, we introduce the notation $\partial\mathcal{A}_{m,n}$ ($1\leq m<d$)
for the $n$-th $\left(d-m\right)$-dimensional relopen face of $\mathcal{A}$
(from here on an open $\left(d-m\right)$-face), while $\partial\mathcal{A}_{d,n}$
is the $n$-th closed $0$-face. The closed $0$-faces are the vertex
of the polytope. The face-based partition of $\partial\mathcal{A}$
is $\left\{ \partial\mathcal{A}_{1,1},\ldots,\partial\mathcal{A}_{m,n},\ldots,\partial\mathcal{A}_{d,n'},\ldots\right\} $
with $\partial\mathcal{A}_{m,i}\cap\partial\mathcal{A}_{m',j}=\textrm{Ø}$
for every $m,i\neq m',j$. Note that $\partial\mathcal{A}_{m,n}$
is $\left[\partial\mathcal{A}\right]_{m,n}$, i.e., the $\left(m,n\right)$-element
of $\partial\mathcal{A}$. Besides, we define 
\begin{equation}
\partial\mathcal{A}_{m}\equiv\bigcup_{n}\partial\mathcal{A}_{m,n}\:,
\end{equation}
as the set of points in $\partial\mathcal{A}$ that lies in some of
the $\left(d-m\right)$-faces of $\mathcal{A}$, with $\left\{ \partial\mathcal{A}_{m,1},\ldots,\partial\mathcal{A}_{m,n},\ldots\right\} $
a partition of $\partial\mathcal{A}_{m}$. Furthermore, 
\begin{equation}
\partial\mathcal{A}=\bigcup_{m}\partial\mathcal{A}_{m}\:,
\end{equation}
with $m$ going from $1$ to $d$, where $\left\{ \partial\mathcal{A}_{1},\ldots,\partial\mathcal{A}_{d}\right\} $
is a partition of $\partial\mathcal{A}$. Given any open face, the
corresponding closed face is its closure. Furthermore, given any closed
$k$-face with $k>0$, its $\textrm{relint}$ is the corresponding
open $k$-face (an open face relative to its affine hull). Some
$k$-faces are designed by their names: a $3$-face is a cell (polyhedron),
a $2$-face is a facet (polygon) and a $1$-face is an edge (line).

The next step is to demonstrate that every CC SB polytope is face-decomposable.
To find the $\left(d-1\right)$-faces of a CC SB polytope one can
exploit that locally they are equal to the $\left(d-1\right)$-faces
of a convex polytope. Let $\mathcal{A}\subseteq\mathbb{P}^{d}$, $\mathbf{r}\in\partial\mathcal{A}$
belongs to an open $\left(d-1\right)$-face of $\mathcal{A}$ if no
matter how small is $\lambda$ with $\lambda\in\mathbb{R}_{>0}$ the
set $\textrm{int}\left[U\left(\mathbf{r},\lambda\right)\cap\mathcal{A}\right]$
is the interior of a half-ball (a ball cut by a hyperplane through
its center). In this case, $\textrm{relint}\left[U\left(\mathbf{r},\lambda\right)\cap\partial\mathcal{A}\right]\subset\partial\mathcal{A}_{1}$
belongs to a given open $\left(d-1\right)$-face and we introduce
the tangent affine space $\textrm{aff}\left(\mathbf{r},\mathcal{\partial A}\right)=\textrm{aff}\left[U\left(\mathbf{r},\lambda\right)\cap\mathcal{\partial A}\right]$
of this face. Let $\mathbf{r},\mathbf{s}\in\partial\mathcal{A}_{1}$
with $\textrm{aff}\left(\mathbf{r},\mathcal{\partial A}\right)=\textrm{aff}\left(\mathbf{s},\mathcal{\partial A}\right)$
and assume that exist a path $C\left(\mathbf{r},\mathbf{s}\right)\subset\mathcal{\partial A}_{1}$
that connects $\mathbf{r}$ and $\mathbf{s}$ such that for every
$\mathbf{x}\in C\left(\mathbf{r},\mathbf{s}\right)$ $\textrm{aff}\left(\mathbf{x},\mathcal{\partial A}\right)=\textrm{aff}\left(\mathbf{r},\mathcal{\partial A}\right)$,
then $\mathbf{r}$, $\mathbf{s}$ and $C\left(\mathbf{r},\mathbf{s}\right)$
are in the same open $\left(d-1\right)$-face of $\mathcal{A}$. Furthermore,
the connectedness-based partition of $\partial\mathcal{A}_{1}$ is
the set of open $\left(d-1\right)$-faces of $\mathcal{A}$, i.e.
$\left\{ \partial\mathcal{A}_{1,1},\cdots,\partial\mathcal{A}_{1,n},\cdots\right\} $.
Given that $\partial\mathcal{A}=\cup_{n}\textrm{cl}\left(\partial\mathcal{A}_{1,n}\right)$,
thus every $\mathbf{r}\in\partial\mathcal{A}\setminus\partial\mathcal{A}_{1}$
belongs to the relative boundary of two $\left(d-1\right)$-faces
(may be more than two). Let $\partial\mathcal{A}_{1,i}$ and $\partial\mathcal{A}_{1,j}$
be two different open $\left(d-1\right)$-faces of $\mathcal{A}$
with $\textrm{cl}\left(\partial\mathcal{A}_{1,i}\right)\cap\textrm{cl}\left(\partial\mathcal{A}_{1,j}\right)\neq\emptyset$,
then $\textrm{cl}\left(\partial\mathcal{A}_{1,i}\right)\cap\textrm{cl}\left(\partial\mathcal{A}_{1,j}\right)$
is a closed $m$-face of $\mathcal{A}$ with $m=\textrm{dim}\left[\textrm{cl}\left(\partial\mathcal{A}_{1,i}\right)\cap\textrm{cl}\left(\partial\mathcal{A}_{1,j}\right)\right]$.
By analyzing each pair of mutually intersecting closed $\left(d-1\right)$-faces
one find each of the remaining $\partial\mathcal{A}_{m,n}$ and $\partial\mathcal{A}_{m}$
elements with $1<m\leq d$. Now, given $\mathcal{A}\in\mathbb{P}^{d}$
we can obtain a non-closed SB polytope by removing from $\mathcal{A}$
one or several of its faces. However, for PW purposes the polytope
represents the integration domain of a well behaved function. Given
that the value of the integral is not modified by removing one or
several faces, we will only consider the case of closed $\mathcal{A}$.

Let $\mathcal{A}\in\mathbb{P}^{d}$, we say that the two elements
$\partial\mathcal{A}_{m,i}$ and $\partial\mathcal{A}_{m',j}$ of
$\partial\mathcal{A}$ are neighbors if $\partial\mathcal{A}_{m,i}\subseteq\textrm{cl}\left(\partial\mathcal{A}_{m',j}\right)$
or $\partial\mathcal{A}_{m',j}\subseteq\textrm{cl}\left(\partial\mathcal{A}_{m,i}\right)$,
for $m,i\neq m',j$. On the other hand, they are adjacent if $\textrm{cl}\left(\partial\mathcal{A}_{m,i}\right)\cap\textrm{cl}\left(\partial\mathcal{A}_{m',j}\right)\neq\emptyset$.
It is simple to verify that if two elements are neighbors they are
adjacent, and the distance between two adjacent elements is zero.
We define the characteristic length of the polytope $\mathcal{A}$
as 
\begin{equation}
\mathfrak{L}\left(\mathcal{A}\right)=\min\left\{ \textrm{d}\left[\partial\mathcal{A}_{m,i},\partial\mathcal{A}_{m',j}\right]\neq0\right\} \:,\label{eq:LdeA}
\end{equation}
i.e. the minimum distance between two non-adjacent elements of $\partial\mathcal{A}$.
It is clear that if $\mathcal{A}\in\mathbb{P}^{d}$, then $\exists$
$\varepsilon\in\mathbb{R}_{>0}$ such that $\mathfrak{L(}\mathcal{A})>\varepsilon$.

In the following lines we state the general proposition for positive
integer values of $d$. The cases of $d=1,2,3$ are demonstrated each
one in a separated proof. The remaining cases that concern every positive
integer value $d\geq4$ are not demonstrated here and thus, the proposition
is for these cases a conjecture. We note that Theorem 2 and its the
conjectured generalization to $\mathbb{R}^{d}$ given in the following
Eq. (\ref{eq:IntEd}) strongly resemble the combination of Hadwiger's
characterization theorem and The general kinematic formula (see Theorems
9.1.1 and 10.3.1 in pp. 118 and 153 of \cite{Klain1991}, respectively)
and transpires analogies with several results of integral geometry.
These points will be discussed at the end of Sec. \ref{sec:Conclu}.

\medskip{}

\label{Th2d}General conjecture (cases d$\geqslant1$): Consider \foreignlanguage{english}{$\mathcal{A}\in\mathbb{P}^{d}$}
and let $G(\mathcal{A},\mathbf{r}):\mathbb{P}^{d}\times\mathbb{R}^{d}\rightarrow\mathbb{R}$
be a well behaved function in $\mathbf{r}\in\textrm{cl}\left(\mathcal{A}\right)$
(with fixed $\mathcal{A}$), and a local function with range $\varsigma\in\mathbb{R}_{>0}$
invariant under rigid transformations with $2\varsigma<\mathfrak{L}\left(\mathcal{A}\right)$
{[}for fixed $\mathcal{A}\in\mathbb{P}^{d}$ function $G(\mathbf{r},\mathcal{A})$
is simply $G(\mathbf{r}):\mathcal{A}\rightarrow\mathbb{R}${]}. Then
\begin{eqnarray}
t & = & \int_{\mathcal{A}}G(\mathbf{r})d\mathbf{r}\nonumber \\
 & = & \chi_{_{d}}c_{0}+\chi_{_{d-1}}c_{1}+\sum_{m=2}^{d}\,\sum_{n\: m\textrm{-elem}}\chi_{_{d-m,n}}c_{m,n}\:,\label{eq:IntEd}
\end{eqnarray}
where $\chi_{_{d}}$ is the $d$-dimensional measure (Lebesgue measure)
of $\mathcal{A}$ (i.e. its volume $V$), $\chi_{_{d-1}}$ is the
$\left(d-1\right)$-dimensional measure of $\partial\mathcal{A}$,
and $\chi_{_{m,n}}$ is the $m$-dimensional measure of the $n$-th
$m$-element of $\partial\mathcal{A}$ (in particular $\chi_{_{0,n}}=1$).
Besides, $c_{0}$ and $c_{1}$ are constant coefficients which are
independent of the size and shape of $\mathcal{A}$, while $c_{m,n}$
is a function of the angles that define the geometry of $\mathcal{A}$
near its $n$-th $m$-element $\partial\mathcal{A}_{d-m,n}$ far away
from its relative boundary.

\medskip{}

\label{Th2}Theorem 2 (case d=3): For the case $d=3$ the Eq. (\ref{eq:IntEd})
is
\begin{eqnarray}
t & = & \int_{\mathcal{A}}G(\mathbf{r})d\mathbf{r}\nonumber \\
 & = & Vc_{0}+Ac_{1}+\sum_{n\textrm{ edges}}L_{n}c_{2,n}+\sum_{n\textrm{ vertex}}c_{3,n}\:,\label{eq:IntE}
\end{eqnarray}
where $V$ is the volume of $\mathcal{A}$, $A$ is the surface area
of $\partial\mathcal{A}$, and $L_{n}$ is the length of the $n$-th
edge. Besides, $c_{0}$ and $c_{1}$ are constant coefficients which
are independent of the size and shape of $\mathcal{A}$, while $c_{2,n}$
is a function of the dihedral angle in the $n$-th edge, and $c_{3,n}$
is a function of the set of dihedral angles between the adjacent planes
that converge to the $n$-th vertex.

\begin{figure}
\centering{}\includegraphics[width=4cm]{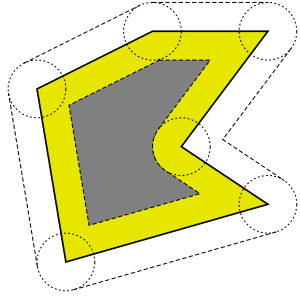}\caption{Picture describing the partition for $\mathcal{A}$ in its principal
and skin parts, $\mathcal{A}_{o}$ and $\mathcal{A}_{\textrm{sk}}$,
respectively.\label{fig:Apartition}}
\end{figure}

\label{ProofTh2}Proof (case d=3): Consider the set $\mathcal{A}\in\mathbb{P}$
and introduce a partition of $\mathcal{A}$ given by $\{\mathcal{A}_{o},\mathcal{A}_{\textrm{sk}}\}$
with the principal part $\mathcal{A}_{o}=\{\mathbf{r}\in\mathcal{A}/d\left(\mathbf{r},\partial\mathcal{A}\right)\geq\varsigma\}$
(the inner parallel body of $\mathcal{A}$) and the boundary or skin
part $\mathcal{A}_{\textrm{sk}}=\mathcal{A}\setminus\mathcal{A}_{o}$
characterized by a thickness $\varsigma$. We have 
\begin{equation}
\mathcal{A}=\mathcal{A}_{o}\cup\,\mathcal{A}_{\textrm{sk}}\:.\label{eq:APart}
\end{equation}
Here $\mathcal{A}_{o}$ is the region where $G(\mathbf{r})$ takes
a constant value. This partition may be obtained by wrapping $\partial\mathcal{A}$
through sliding the center of a ball of radius $\varsigma$ on $\partial\mathcal{A}$.
Let $w\left(\partial\mathcal{A},\varsigma\right)$ be the set of points
that are inside of any of this balls, then $\mathcal{A}_{o}=\mathcal{A}\setminus w\left(\partial\mathcal{A},\varsigma\right)$.
Besides, it may also be obtained in terms of a Minkowski sum. Let
$U$ be the unit ball centered at the origin, then the Minkowski sum
$\mathcal{A}^{c}+\varsigma U$ is the outer parallel body of $\mathcal{A}^{c}$
\cite{Schneider2008} and $\mathcal{A}_{o}=\mathcal{A}\setminus\left(\mathcal{A}^{c}+\varsigma U\right)$.
In Fig. \ref{fig:Apartition} a picture representing this partition
for $\mathcal{A}$ is shown. There, one can observe $\partial\mathcal{A}$
on continuous line and several balls of radius $\varsigma$ (in dotted
lines) corresponding to the wrapping procedure. For the case of $\mathcal{A}$
being the shaded region (both the darker and brighter ones), the darker
shaded region represents $\mathcal{A}_{o}$ while the brighter shaded
one corresponds to $\mathcal{A}_{\textrm{sk}}$. On the contrary,
if $\mathcal{A}$ is the non-shaded or white region then $\mathcal{A}_{\textrm{sk}}$
corresponds to the white region between $\partial\mathcal{A}$ and
the dashed line while the rest of the white region represents $\mathcal{A}_{o}$.
For both cases a dashed line separates regions $\mathcal{A}_{o}$
and $\mathcal{A}_{\textrm{sk}}$.
 The integral in Eq. (\ref{eq:IntE}) gives
\begin{equation}
\int_{\mathcal{A}}G(\mathbf{r})d\mathbf{r}=Vc_{0}+\int_{\mathcal{A}_{\textrm{sk}}}g(\mathbf{r})d\mathbf{r}\:,\label{eq:IntE01}
\end{equation}
with $g(\mathbf{r})=G(\mathbf{r})-c_{0}$ and $c_{0}=G(\mathbf{r})$
for all $\mathbf{r}\in\mathcal{A}_{o}$. Naturally, $g(\mathbf{r}):\mathcal{A}\rightarrow\mathbb{R}$
is a bounded and local function with range $\varsigma$ invariant
under rigid transformations which implies that $g(\mathbf{r})=g(U(\mathbf{r},\varsigma)\cap\mathcal{A})$
and its value is independent of the position and orientation that
the set $U(\mathbf{r},\varsigma)\cap\mathcal{A}$ takes on the space,
$g(\mathbf{r})$ only depends on the shape of $U(\mathbf{r},\varsigma)\cap\mathcal{A}$.
Besides if $\mathbf{r}\notin\mathcal{A}_{\textrm{sk}}$ then $g(\mathbf{r})=0$.

Note that in the context of $d=3$ a planar face is simply a face.
Consider the set $\mathcal{A}_{\textrm{sk}}$ partitioned in terms
of $\left\{ \mathcal{A}_{1},\mathcal{A}_{2},\mathcal{A}_{3}\right\} $,
\begin{equation}
\mathcal{A}_{\textrm{sk}}=\mathcal{A}_{1}\cup\mathcal{A}_{2}\cup\mathcal{A}_{3}\:.\label{eq:AskPart}
\end{equation}
Here, $\mathcal{A}_{1}$ is the surface-type region $\mathcal{A}_{1}=\left\{ \mathbf{r}\in\mathcal{A}_{\textrm{sk}}/U(\mathbf{r},\varsigma)\cap\partial\mathcal{A}_{1,n}\neq\emptyset\textrm{ for only one value of }n\textrm{ face}\right\} $
where each $\mathbf{r}\in\mathcal{A}_{1}$ is related to a unique
face. Let us define $\mathcal{A}_{1,n}=\left\{ \mathbf{r}\in\mathcal{A}_{1}/U(\mathbf{r},\varsigma)\cap\partial\mathcal{A}_{1,n}\neq\emptyset\right.$
$\left.\textrm{ for the }n\textrm{-th face}\right\} $ thus $\{\mathcal{A}_{1,1},\cdots,\mathcal{A}_{1,n},\cdots\}$
is a partition of $\mathcal{A}_{1}$. Since $\mathbf{r}\in\mathcal{A}_{1}$
then $\mathbf{r}\in\mathcal{A}_{1,n}$ for some $n$ face. Hence,
the shape of $U(\mathbf{r},\varsigma)\cap\mathcal{A}$ is determined
by the one dimensional position variable $\boldsymbol{r}^{(1)}=\textrm{d}(\mathbf{r},\partial\mathcal{A}_{1,n})=z$,
and then, $g(\mathbf{r})=g_{1}(\mathbf{r})=g_{1}\left(\boldsymbol{r}^{(1)}\right)=g_{1}\left(z\right)$.
In Eq. (\ref{eq:AskPart}) $\mathcal{A}_{2}$ is the edges-type region
$\mathcal{A}_{2}=\left\{ \mathbf{r}\in\mathcal{A}_{\textrm{sk}}/U(\mathbf{r},\varsigma)\cap\partial\mathcal{A}_{1,m}\neq\emptyset\textrm{ for two and only two}\right.$
$\left.\textrm{values of }m\textrm{ faces}\right\} $ where each $\mathbf{r}\in\mathcal{A}_{2}$
is related to a unique pair of faces. Let $\mathbf{r}\in\mathcal{A}_{2}$
and $m_{1},m_{2}$ be such that $U(\mathbf{r},\varsigma)\cap\partial\mathcal{A}_{1,m_{1}}\neq\emptyset$
and $U(\mathbf{r},\varsigma)\cap\partial\mathcal{A}_{1,m_{2}}\neq\emptyset$,
given that $0<2\varsigma<\mathfrak{L(}\mathcal{A})$ the pair of faces
$m_{1},m_{2}$ converge to a common edge. Let us define $\mathcal{A}_{2,n}=\left\{ \mathbf{r}\in\mathcal{A}_{2}/\right.$
$\left.U(\mathbf{r},\varsigma)\cap\partial\mathcal{A}_{1,m}\neq\emptyset\right.$
$\left.\textrm{for the two values of }m\textrm{ faces that join in the }n\textrm{-th edge}\right\} $
thus $\{\mathcal{A}_{2,1},\cdots,\mathcal{A}_{2,n},\cdots\}$ is a
partition of $\mathcal{A}_{2}$. Since $\mathbf{r}\in\mathcal{A}_{2}$
then $\mathbf{r}\in\mathcal{A}_{2,n}$ for only one edge (the $n$-th).
Hence, the shape of $U(\mathbf{r},\varsigma)\cap\mathcal{A}$ is determined
by both, the dihedral angle and the two-dimensional vector $\boldsymbol{r}^{(2)}$
that lies in a plane orthogonal to $\partial\mathcal{A}_{2,n}$ and
goes from $\partial\mathcal{A}_{2,n}$ to $\mathbf{r}$, and thus,
$g(\mathbf{r})=g_{2}(\mathbf{r})=g_{2}(\boldsymbol{r}^{(2)})$. $\mathcal{A}_{3}$
is the vertex-type region $\mathcal{A}_{3}=\bigl\{\mathbf{r}\in\mathcal{A}_{\textrm{sk}}/U(\mathbf{r},\varsigma)\cap\partial\mathcal{A}_{1,m}\neq\emptyset$
$\textrm{for three or more values of }m\textrm{ face}\bigr\}$ where
each $\mathbf{r}$ is related to a unique set of faces. Let $\mathbf{r}\in\mathcal{A}_{3}$
and $\{m_{1},m_{2},m_{3},...\}$ be such that $U(\mathbf{r},\varsigma)\cap\partial\mathcal{A}_{1,m_{1}}\neq\emptyset$,
$U(\mathbf{r},\varsigma)\cap\partial\mathcal{A}_{1,m_{2}}\neq\emptyset$,
$U(\mathbf{r},\varsigma)\cap\partial\mathcal{A}_{1,m_{3}}\neq\emptyset$
..., given that $0<2\varsigma<\mathfrak{L(}\mathcal{A})$ the set
of faces $\{m_{1},m_{2},m_{3},...\}$ converge to one vertex. Let
us define $\mathcal{A}_{3,n}=\left\{ \mathbf{r}\in\mathcal{A}_{3}/\right.$$\left.U(\mathbf{r},\varsigma)\cap\partial\mathcal{A}_{1,m}\neq\emptyset\right.$
$\left.\textrm{for the }i\textrm{ values (with }i\geq3\textrm{) of }m\textrm{ faces that join in the }n\textrm{ vertex}\right\} $
thus $\{\mathcal{A}_{3,1},\cdots,\mathcal{A}_{3,n},\cdots\}$ is a
partition of $\mathcal{A}_{3}$. Since $\mathbf{r}\in\mathcal{A}_{3}$
then $\mathbf{r}\in\mathcal{A}_{3,n}$ for some $n$ vertex. Hence,
the shape of $U(\mathbf{r},\varsigma)\cap\mathcal{A}$ is determined
by the three-dimensional vector $\boldsymbol{r}^{(3)}$ that goes
from $\partial\mathcal{A}_{3,n}$ to $\mathbf{r}$ (and the angles
that define the geometry of the vertex). Naturally, $g(\mathbf{r})=g_{3}(\mathbf{r})=g_{3}(\boldsymbol{r}^{(3)})$.
Care must be taken about $\partial\mathcal{A}_{m,n}$ {[}the $\left(m,n\right)$
element of $\partial\mathcal{A}${]} that should not be confused with
$\partial\left(\mathcal{A}_{m,n}\right)$ {[}the boundary of the $\left(m,n\right)$
element of the partition of $\mathcal{A}${]} an irrelevant magnitude
in PW.

Using the partition introduced for $\mathcal{A}_{1}$, $\mathcal{A}_{2}$,
$\mathcal{A}_{3}$ and the properties of $g(\mathbf{r})$ we obtain
\begin{equation}
\int_{\mathcal{A}_{\textrm{sk}}}g(\mathbf{r})d\mathbf{r}=\sum_{n\textrm{ faces}}\int_{\mathcal{A}_{1,n}}g_{1}(\mathbf{r})d\mathbf{r}+\sum_{n\textrm{ edges}}\int_{\mathcal{A}_{2,n}}g_{2}(\mathbf{r})d\mathbf{r}+\sum_{n\textrm{ vertex}}\int_{\mathcal{A}_{3,n}}g_{3}(\mathbf{r})d\mathbf{r}\:,\label{eq:IntAsk}
\end{equation}
where the right hand side term is a contribution to each vertex similar
to that appearing in Eq. (\ref{eq:IntE}). Each of the edge's terms
\begin{equation}
\int_{\mathcal{A}_{2,n}}g_{2}(\mathbf{r})d\mathbf{r}\:,\label{eq:IntA2ng2}
\end{equation}
can be treated separately. Let us consider the family of orthogonal
planes to the direction of the $n$-th edge which intersects $\partial\mathcal{A}_{2,n}$.
Each of these planes cuts $\mathcal{A}_{2,n}$ producing a slice or
orthogonal cross section of $\mathcal{A}_{2,n}$. Let $\mathcal{A}_{2,n}^{\perp}$
be the cross section defined by one of such planes that intersects
$\mathcal{A}_{2,n}$ in a region where both endpoints of the $n$-th
edge are away to ensure that the shape of $\mathcal{A}_{2,n}^{\perp}$
does not depend on any arbitrary choice. One can define $\mathcal{B}_{2,n}$
as the righted version of $\mathcal{A}_{2,n}$, which is the right
(generalized) cylinder obtained by translate $\mathcal{A}_{2,n}^{\perp}$
along $\partial\mathcal{A}_{2,n}$, and extend the domain of $g_{2}(\mathbf{r})$
to $\mathcal{B}_{2,n}$. The set $\mathcal{B}_{2,n}\setminus\mathcal{A}_{2,n}$
contains two disconnected regions one around each endpoint of $\partial\mathcal{A}_{2,n}$
(related to a given vertex). Each of these regions is $\left(\mathcal{B}_{2,n}\setminus\mathcal{A}_{2,n}\right)_{n'}$
with $n'$ running over the vertex of $\mathcal{A}$ and $\left(\mathcal{B}_{2,n}\setminus\mathcal{A}_{2,n}\right)_{n'}=\emptyset$
if edge $n$ and vertex $n'$ are not neighbors. Therefore
\begin{equation}
\int_{\mathcal{A}_{2,n}}g_{2}(\mathbf{r})d\mathbf{r}=\int_{\mathcal{B}_{2,n}}g_{2}(\mathbf{r})d\mathbf{r}-\sum_{n'\textrm{vertex}}\int_{\left(\mathcal{B}_{2,n}\setminus\mathcal{A}_{2,n}\right)_{n'}}g_{2}(\mathbf{r})d\mathbf{r}\:,\label{eq:iA2ng2}
\end{equation}
 with
\begin{equation}
\int_{\mathcal{B}_{2,n}}g_{2}(\mathbf{r})d\mathbf{r}=L_{n}\int_{\mathcal{A}_{2,n}^{\perp}}g_{2}(\boldsymbol{r}^{(2)})d\boldsymbol{r}^{(2)}\:,\label{eq:iB2ng2}
\end{equation}
where $\boldsymbol{r}^{(2)}$ is the two-dimensional coordinate of
a point in $\mathcal{A}_{2,n}^{\perp}$ and the last integral is a
function of the dihedral angle at the $n$-th edge. Hence, for the
central term in the right-hand side of Eq. (\ref{eq:IntAsk}) we found
\begin{equation}
\sum_{n\textrm{ edges}}\int_{\mathcal{A}_{2,n}}g_{2}(\mathbf{r})d\mathbf{r}=\sum_{n\textrm{ edges}}L_{n}\int_{\mathcal{A}_{2,n}^{\perp}}g_{2}(\boldsymbol{r}^{(2)})d\boldsymbol{r}^{(2)}-\sum_{n'\textrm{vertex}}\sum_{n\textrm{ edge}}\int_{\left(\mathcal{B}_{2,n}\setminus\mathcal{A}_{2,n}\right)_{n'}}g_{2}(\boldsymbol{r}^{(2)})d\boldsymbol{r}^{(3)}\:.\label{eq:siA2ng2}
\end{equation}
Now, to analyze each of the face terms in Eq. (\ref{eq:IntAsk})
\begin{equation}
\int_{\mathcal{A}_{1,n}}g_{1}(\mathbf{r})d\mathbf{r}\:,\label{eq:IntA1ng1}
\end{equation}
let us define $\mathcal{B}_{1,n}$ as the righted version of $\mathcal{A}_{1,n}$,
which is the right (generalized) cylinder obtained by translate the
face $\partial\mathcal{A}_{1,n}$ along $\varsigma\hat{z}_{n}$ being
$\hat{z}_{n}$ the inner (to $\mathcal{A}$) normal unit vector, note
that $\mathcal{A}_{1,n}\subset\mathcal{B}_{1,n}$. One can extend
the domain of $g_{1}(\mathbf{r})$ to all the semi-space that contains
$\mathcal{B}_{1,n}$ and includes $\partial\mathcal{A}_{1,n}$ in
its boundary plane to obtain
\begin{equation}
\int_{\mathcal{A}_{1,n}}g_{1}(\mathbf{r})d\mathbf{r}=\int_{\mathcal{B}_{1,n}}g_{1}(\mathbf{r})d\mathbf{r}-\int_{\mathcal{B}_{1,n}\setminus\mathcal{A}_{1,n}}g_{1}(\mathbf{r})d\mathbf{r}\:,\label{eq:iA1ng1}
\end{equation}
 with
\begin{equation}
\int_{\mathcal{B}_{1,n}}g_{1}(\mathbf{r})d\mathbf{r}=A_{n}\int_{0}^{\varsigma}g_{1}(z)dz\:,\label{eq:iB1ng1}
\end{equation}
where the last integral is independent of the involved face. Besides,
$\mathcal{B}_{1,n}\setminus\mathcal{A}_{1,n}$ is a set of points
distributed in the neighborhood of the relative boundary of $\mathcal{A}_{1,n}$
i.e. $\textrm{rel}\partial\left(\mathcal{A}_{1,n}\right)=\cup_{n'}\partial\mathcal{A}_{2,n'}\cup_{n''}\partial\mathcal{A}_{3,n''}$
$ $ with $n'$ ($n^{\prime\prime}$) running over the edges (vertex)
that are neighbors of $\partial\mathcal{A}_{1,n}$. We make a partition
of $\mathcal{B}_{1,n}\setminus\mathcal{A}_{1,n}$ in regions corresponding
to edges and vertex $\{\ldots,\mathcal{B}_{2,nn'}^{\prime},\ldots,\mathcal{B}_{3,nn''}^{\prime},\ldots\}$
\begin{equation}
\mathcal{B}_{1,n}\setminus\mathcal{A}_{1,n}=\bigcup_{n'\textrm{edges}}\mathcal{B}_{2,nn'}^{\prime}\bigcup_{n''\textrm{vertex}}\mathcal{B}_{3,nn''}^{\prime}\:,\label{eq:BA1part}
\end{equation}
therefore
\begin{equation}
\int_{\mathcal{B}_{1,n}\setminus\mathcal{A}_{1,n}}g_{1}(\mathbf{r})d\mathbf{r}=\sum_{n'\textrm{edges}}\int_{\mathcal{B}_{2,nn'}^{\prime}}g_{1}(\mathbf{r})d\mathbf{r}+\sum_{n''\textrm{vertex}}\int_{\mathcal{B}_{3,nn''}^{\prime}}g_{1}(\mathbf{r})d\mathbf{r}\:,\label{eq:iBA1ng1}
\end{equation}
where again $n'$ ($n''$) runs over the edges (vertex) that are neighbors
of the $n$ face. This partition is a simple extension of $\mathcal{A}_{2,n'}$
and $\mathcal{A}_{3,n''}$, and it is warranted because the boundaries
between $\mathcal{A}_{1,n}$ $\mathcal{A}_{2,n'}$ and $\mathcal{A}_{3,n''}$
are built by pieces of planes, cylinders and spheres that can be trivially
extended to such regions $\mathcal{B}_{1,n}\setminus\mathcal{A}_{1,n}\nsubseteq\mathcal{A}$.
For the $n'$-th edge we consider the family of orthogonal planes
that cut $\mathcal{B}_{2,nn'}^{\prime}$ each plane producing a slice
or orthogonal cross section. Let $\left(\mathcal{B}_{1,n}\setminus\mathcal{A}_{1,n}\right)_{n'}^{\perp}$
be one of such cross section that cut $\partial\mathcal{A}_{2,n'}$
in a region where both endpoints of the $n'$ edge are away to ensure
that the cross section does not depend on any arbitrary choice. For
each $n'$ edge we define $\mathcal{B}_{2,nn'}^{\prime\prime}$ as
the right (generalized) cylinder obtained by moving $\left(\mathcal{B}_{1,n}\setminus\mathcal{A}_{1,n}\right)_{n'}^{\perp}$
along $\partial\mathcal{A}_{2,n'}$. We note that $\mathcal{B}_{2,nn'}^{\prime\prime}\setminus\mathcal{B}_{2,nn'}^{\prime}$
contains two disjoint parts one around each endpoint of $\partial\mathcal{A}_{2,n'}$
which we assign to the corresponding vertex. Each of these contributions
is $\left(\mathcal{B}_{2,nn'}^{\prime\prime}\setminus\mathcal{B}_{2,nn'}^{\prime}\right)_{n''}$
with $n'$ running over the vertex and $\left(\mathcal{B}_{2,nn'}^{\prime\prime}\setminus\mathcal{B}_{2,nn'}^{\prime}\right)_{n''}=\emptyset$
if face $n$, edge $n'$ and vertex $n''$ are not all neighbors (taken
in pairs). Therefore, for a given $n$ face and $n'$ edge we have
\begin{equation}
\int_{\mathcal{B}_{2,nn'}^{\prime}}g_{1}(\mathbf{r})d\mathbf{r}=\int_{\mathcal{B}_{2,nn'}^{\prime\prime}}g_{1}(\mathbf{r})d\mathbf{r}-\sum_{n''\textrm{vertex}}\int_{\left(\mathcal{B}_{2,nn'}^{\prime\prime}\setminus\mathcal{B}_{2,nn'}^{\prime}\right)_{n''}}g_{1}(\mathbf{r})d\mathbf{r}\:,\label{eq:iBp2nng1}
\end{equation}
 with
\begin{equation}
\int_{\mathcal{B}_{2,nn'}^{\prime\prime}}g_{1}(\mathbf{r})d\mathbf{r}=L_{n'}\int_{\left(\mathcal{B}_{1,n}\setminus\mathcal{A}_{1,n}\right)_{n'}^{\perp}}g_{1}(z)d\boldsymbol{r}^{(2)}\:.\label{eq:iBpp2nng1}
\end{equation}
One can verify that the edge contribution in Eq. (\ref{eq:iBpp2nng1})
for each of the two faces that meet at the $n'$-th edge is the same.
By adding the contribution of all the faces, i.e. joining results
from Eq. (\ref{eq:iA1ng1}) to Eq. (\ref{eq:iBpp2nng1}), one found
\begin{eqnarray}
\sum_{n\textrm{ faces}}\int_{\mathcal{A}_{1,n}}g_{1}(\mathbf{r})d\mathbf{r} & = & A\int_{0}^{\varsigma}g_{1}(z)dz-2\sum_{n'\textrm{edges}}L_{n'}\int_{\left(\mathcal{B}_{1,m}\setminus\mathcal{A}_{1,m}\right)_{n'}^{\perp}}g_{1}(z)d\boldsymbol{r}^{(2)}\:\nonumber \\
 &  & +\sum_{n''\textrm{vert}}\sum_{n\textrm{ face}}\left[\sum_{n'\textrm{edge}}\int_{\left(\mathcal{B}_{2,nn'}^{\prime\prime}\setminus\mathcal{B}_{2,nn'}^{\prime}\right)_{n''}}g_{1}(z)d\boldsymbol{r}^{(3)}-\int_{\mathcal{B}_{3,nn''}^{\prime}}g_{1}(z)d\mathbf{r}^{(3)}\right]\:,\label{eq:siA1ng1}
\end{eqnarray}
where the $m$ label (which is not an index) corresponds to any of
both faces that meet at the $n'$ edge. Putting all together, one
obtain
\begin{equation}
t=Vc_{0}+Ac_{1}+\sum_{n\textrm{ edges}}L_{n}c_{2,n}+\sum_{n\textrm{ vertex}}c_{3,n}\:,\label{eq:IntT}
\end{equation}
with the coefficients 
\begin{equation}
c_{0}=G(\mathbf{r})\textrm{ for all }\mathbf{r}\in\mathcal{A}_{o}\:,\label{eq:defc0}
\end{equation}
\begin{equation}
c_{1}=\int_{0}^{\varsigma}g_{1}(z)dz\:,\label{eq:defc1}
\end{equation}
\begin{equation}
c_{2,n}=\int_{\mathcal{A}_{2,n}^{\perp}}g_{2}(\boldsymbol{r}^{(2)})d\boldsymbol{r}^{(2)}-2\int_{\left(\mathcal{B}_{1,m}\setminus\mathcal{A}_{1,m}\right)_{n}^{\perp}}g_{1}(z)d\boldsymbol{r}^{(2)}\:,\label{eq:defc2}
\end{equation}
\begin{eqnarray}
c_{3,n} & = & \int_{\mathcal{A}_{3,n}}g_{3}(\boldsymbol{r}^{(3)})d\boldsymbol{r}^{(3)}-\sum_{n'\textrm{edges}}\int_{\left(\mathcal{B}_{2,n'}\setminus\mathcal{A}_{2,n'}\right)_{n}}g_{2}(\boldsymbol{r}^{(2)})d\boldsymbol{r}^{(3)}\:\nonumber \\
 &  & +\sum_{n''\textrm{faces}}\left[\sum_{n'\textrm{edges}}\int_{\left(\mathcal{B}_{2,n''n'}^{\prime\prime}\setminus\mathcal{B}_{2,n''n'}^{\prime}\right)_{n}}g_{1}(z)d\boldsymbol{r}^{(3)}-\int_{\mathcal{B}_{3,n''n}^{\prime}}g_{1}(z)d\boldsymbol{r}^{(3)}\right]\:,\label{eq:defc3}
\end{eqnarray}
and $g(\mathbf{r})=G(\mathbf{r})-c_{0}$.\hfill{}$\blacksquare$

\medskip{}

\label{Th2-1}Theorem 2 (case d=2): For the case $d=2$ Eq. (\ref{eq:IntEd})
reads
\begin{eqnarray}
t & = & \int_{\mathcal{A}}G(\mathbf{r})d\mathbf{r}\nonumber \\
 & = & Ac_{0}+Lc_{1}+\sum_{n\textrm{ vertex}}c_{2,n}\:,\label{eq:IntE-1}
\end{eqnarray}
where $A$ is the surface area of $\mathcal{A}$ and $L$ is the length
of its perimeter $\partial\mathcal{A}$. Besides, $c_{0}$ and $c_{1}$
are constant coefficients which are independent of the size and shape
of $\mathcal{A}$, while $c_{2,n}$ is a function of the angle in
the $n$-th vertex.

\label{ProofTh2-1}Proof (case d=2): For brevity, we only present
such parts of the demonstration that differs from the case $d=3$.
Introduce a partition of $\mathcal{A}$ given by $\{\mathcal{A}_{o},\mathcal{A}_{\textrm{sk}}\}$
to obtain
\begin{equation}
\int_{\mathcal{A}}G(\mathbf{r})d\mathbf{r}=Ac_{0}+\int_{\mathcal{A}_{\textrm{sk}}}g(\mathbf{r})d\mathbf{r}\:.\label{eq:IntE01-1}
\end{equation}
Consider the set $\mathcal{A}_{\textrm{sk}}$ partitioned in terms
of $\left\{ \mathcal{A}_{1},\mathcal{A}_{2}\right\} $, 
\begin{equation}
\mathcal{A}_{\textrm{sk}}=\mathcal{A}_{1}\cup\mathcal{A}_{2}\:.\label{eq:AskPart-1}
\end{equation}
$\mathcal{A}_{1}$ is the line-type region $\mathcal{A}_{1}=\left\{ \mathbf{r}\in\mathcal{A}_{\textrm{sk}}/U(\mathbf{r},\varsigma)\cap\partial\mathcal{A}_{1,n}\neq\emptyset\textrm{ for only one }n\textrm{ side}\right\} $
where each $\mathbf{r}\in\mathcal{A}_{1}$ is related to a unique
side. Let us define $\mathcal{A}_{1,n}=\left\{ \mathbf{r}\in\mathcal{A}_{1}/U(\mathbf{r},\varsigma)\cap\partial\mathcal{A}_{1,n}\neq\emptyset\right.$
$\left.\textrm{ for the }n\textrm{-th side}\right\} $ thus $\{\mathcal{A}_{1,1},\cdots,\mathcal{A}_{1,n},\cdots\}$
is a partition of $\mathcal{A}_{1}$. Since $\mathbf{r}\in\mathcal{A}_{1}$
then $\mathbf{r}\in\mathcal{A}_{1,n}$ for only one side (the $n$-th).
Hence, the shape of $U(\mathbf{r},\varsigma)\cap\mathcal{A}$ is determined
by the one dimensional position variable $\boldsymbol{r}^{(1)}=\textrm{d}(\mathbf{r},\partial\mathcal{A}_{1,n})=z$,
and then, $g(\mathbf{r})=g_{1}(\mathbf{r})=g_{1}\left(\boldsymbol{r}^{(1)}\right)=g_{1}\left(z\right)$.
$\mathcal{A}_{2}$ is the vertex-type region $\mathcal{A}_{2}=\left\{ \mathbf{r}\in\mathcal{A}_{\textrm{sk}}/U(\mathbf{r},\varsigma)\cap\partial\mathcal{A}_{1,m}\neq\emptyset\textrm{ for two and only two}\right.$
$\left.\textrm{values of }m\textrm{ side}\right\} $ where each $\mathbf{r}\in\mathcal{A}_{2}$
is related to a unique pair of sides. Let $\mathbf{r}\in\mathcal{A}_{2}$
and $m_{1},m_{2}$ be such that $U(\mathbf{r},\varsigma)\cap\partial\mathcal{A}_{1,m_{1}}\neq\emptyset$
and $U(\mathbf{r},\varsigma)\cap\partial\mathcal{A}_{1,m_{2}}\neq\emptyset$,
given that $0<2\varsigma<\mathfrak{L(}\mathcal{A})$ the pair of sides
$m_{1},m_{2}$ converge to a common vertex. Let us define $\mathcal{A}_{2,n}=\left\{ \mathbf{r}\in\mathcal{A}_{2}/\right.$$\left.U(\mathbf{r},\varsigma)\cap\partial\mathcal{A}_{1,m}\neq\emptyset\right.$$\left.\textrm{for the two values of }m\textrm{ sides that join in the }n\textrm{-th vertex}\right\} $
thus $\{\mathcal{A}_{2,1},\cdots,\mathcal{A}_{2,n},\cdots\}$ is a
partition of $\mathcal{A}_{2}$. Since $\mathbf{r}\in\mathcal{A}_{2}$
then $\mathbf{r}\in\mathcal{A}_{2,n}$ for only one vertex (the $n$-th).
Hence, the shape of $U(\mathbf{r},\varsigma)\cap\mathcal{A}$ is determined
by both, the vertex angle and the vector $\boldsymbol{r}^{(2)}$ that
goes from $\partial\mathcal{A}_{2,n}$ to $\mathbf{r}$, and thus,
$g(\mathbf{r})=g_{2}(\mathbf{r})=g_{2}(\boldsymbol{r}^{(2)})$.

Using the partition introduced for $\mathcal{A}_{1}$, $\mathcal{A}_{2}$
and the properties of $g(\mathbf{r})$ one obtain 
\begin{equation}
\int_{\mathcal{A}_{\textrm{sk}}}g(\mathbf{r})d\mathbf{r}=\sum_{n\textrm{ sides}}\int_{\mathcal{A}_{1,n}}g_{1}(\mathbf{r})d\mathbf{r}+\sum_{n\textrm{ vertex}}\int_{\mathcal{A}_{2,n}}g_{2}(\mathbf{r})d\mathbf{r}\:,\label{eq:IntAsk-1}
\end{equation}
where the right hand side integral is part of the $n$-th vertex contribution
to $t$, similar to that appearing in Eq. (\ref{eq:IntE-1}). To analyze
each of the side terms in Eq. (\ref{eq:AskPart-1})
\begin{equation}
\int_{\mathcal{A}_{1,n}}g_{1}(\mathbf{r})d\mathbf{r}\:,\label{eq:IntA1ng1-1}
\end{equation}
it is convenient to define $\mathcal{B}_{1,n}$ as the righted version
of $\mathcal{A}_{1,n}$, which is the rectangle obtained by translate
the side $\partial\mathcal{A}_{1,n}$ along $\varsigma\hat{z}_{n}$
being $\hat{z}_{n}$ the inner (to $\mathcal{A}$) normal unit vector,
note that $\mathcal{A}_{1,n}\subset\mathcal{B}_{1,n}$. Let us consider
the domain of $g_{1}(\mathbf{r})$ extended to all the semi-plane
that includes $\partial\mathcal{A}_{1,n}$ and contains $\mathcal{B}_{1,n}$.
In terms of $\mathcal{B}_{1,n}$ the Eq. (\ref{eq:IntA1ng1-1}) can
be written as
\begin{equation}
\int_{\mathcal{A}_{1,n}}g_{1}(\mathbf{r})d\mathbf{r}=\int_{\mathcal{B}_{1,n}}g_{1}(\mathbf{r})d\mathbf{r}-\int_{\mathcal{B}_{1,n}\setminus\mathcal{A}_{1,n}}g_{1}(\mathbf{r})d\mathbf{r}\:,\label{eq:iA1ng1-1}
\end{equation}
 with
\begin{equation}
\int_{\mathcal{B}_{1,n}}g_{1}(\mathbf{r})d\mathbf{r}=L_{n}\int_{0}^{\varsigma}g_{1}(z)dz\:,\label{eq:iB1ng1-1}
\end{equation}
where $L_{n}$ is the length of the $n$-th side and the right hand
side integral is independent of the involved side. Furthermore, $\mathcal{B}_{1,n}\setminus\mathcal{A}_{1,n}$
contains two disjoint parts one around each endpoint of $\partial\mathcal{A}_{1,n}$
which one assign to the corresponding vertex. Each of these contributions
is $\left(\mathcal{B}_{1,n}\setminus\mathcal{A}_{1,n}\right)_{n'}$
with $n'$ running over the vertex and $\left(\mathcal{B}_{1,n}\setminus\mathcal{A}_{1,n}\right)_{n'}=\emptyset$
if side $n$ and vertex $n'$ are not neighbors. Adding the contribution
of all the sides, joining results from Eq. (\ref{eq:iA1ng1-1}) to
Eq. (\ref{eq:iB1ng1-1}) and taking into account that each vertex
contributes (with identical contribution) to a pair of sides it is
found
\begin{eqnarray}
\sum_{n\textrm{ sides}}\int_{\mathcal{A}_{1,n}}g_{1}(\mathbf{r})d\mathbf{r} & = & L\int_{0}^{\varsigma}g_{1}(z)dz-2\sum_{n'\textrm{vertex}}\int_{\left(\mathcal{B}_{1,m}\setminus\mathcal{A}_{1,m}\right)_{n'}}g_{1}(z)d\boldsymbol{r}^{(2)}\:\label{eq:siA1ng1-1}
\end{eqnarray}
where the $m$ label corresponds to any of both sides that meet at
the $n'$ vertex (and each side must be considered once). Putting
all together, one obtain
\begin{equation}
t=Ac_{0}+Lc_{1}+\sum_{n\textrm{ vertex}}c_{2,n}\:,\label{eq:IntT-1}
\end{equation}
with the coefficients 
\begin{equation}
c_{0}=G(\mathbf{r})\textrm{ for all }\mathbf{r}\in\mathcal{A}_{o}\:,\label{eq:defc0-1}
\end{equation}
\begin{equation}
c_{1}=\int_{0}^{\varsigma}g_{1}(z)dz\:,\label{eq:defc1-1}
\end{equation}
\begin{equation}
c_{2,n}=\int_{\mathcal{A}_{2,n}}g_{2}(\boldsymbol{r}^{(2)})d\boldsymbol{r}^{(2)}-2\int_{\left(\mathcal{B}_{1,m}\setminus\mathcal{A}_{1,m}\right)_{n}}g_{1}(z)d\boldsymbol{r}^{(2)}\:.\label{eq:defc2-1}
\end{equation}
\hfill{}$\blacksquare$

\medskip{}

\label{Th2-2}Theorem 2 (case d=1): For the case $d=1$ the Eq. (\ref{eq:IntEd})
reads
\begin{eqnarray}
t & = & \int_{\mathcal{A}}G(\mathbf{r})d\mathbf{r}\nonumber \\
 & = & Lc_{0}+2c_{1}\:,\label{eq:IntE-2-1}
\end{eqnarray}
where $L$ is the length of $\mathcal{A}$ while $\partial\mathcal{A}$
is composed by two separated points. Besides, $c_{0}$ and $c_{1}$
are both constant coefficients which are independent of the size $\mathcal{A}$.
Note that, being $\mathcal{A}$ a straight line its shape is unique.
Furthermore, in this case $L=\mathfrak{L(}\mathcal{A})$.

\label{ProofTh2-1-1}Proof (case d=1): Again, we present such parts
of the demonstration that differs from the cases $d=2,3$. Introduce
a partition of $\mathcal{A}$ given by $\{\mathcal{A}_{o},\mathcal{A}_{sk}\}$
to obtain

\begin{equation}
\int_{\mathcal{A}}G(\mathbf{r})d\mathbf{r}=Lc_{0}+\int_{\mathcal{A}_{\textrm{sk}}}g(\mathbf{r})d\mathbf{r}\:.\label{eq:IntE01-2}
\end{equation}
with $g(\mathbf{r})=G(\mathbf{r})-c_{0}$ and $c_{0}=G(\mathbf{r})$
for all $\mathbf{r}\in\mathcal{A}_{o}$. Now, $\mathcal{A}_{\textrm{sk}}=\mathcal{A}_{1}=\left\{ \mathcal{A}_{1,1},\mathcal{A}_{1,2}\right\} $
is a pure vertex-type region. Here, $\mathcal{A}_{1,1}$ and $\mathcal{A}_{1,2}$
are disjoint regions and each one corresponds to a given vertex or
point, being $\partial A_{1,n}\subset\mathcal{A}_{1,n}$. Since both
vertex have the same shape and the shape of $U(\mathbf{r},\varsigma)\cap\mathcal{A}_{1,n}$
is determined by the one dimensional position variable $\boldsymbol{r}^{(1)}=\textrm{d}(\mathbf{r},\partial\mathcal{A}_{1,n})=z$,
then $g(\mathbf{r})=g_{1}(\mathbf{r})=g_{1}\left(\boldsymbol{r}^{(1)}\right)=g_{1}\left(z\right)$.
Thus, one find
\begin{equation}
\int_{\mathcal{A}_{\textrm{sk}}}g(\mathbf{r})d\mathbf{r}=2c_{1}=2\int_{0}^{\varsigma}g_{1}(z)dz\:.\label{eq:IntAsk-2}
\end{equation}
The Eqs. (\ref{eq:IntE01-2}, \ref{eq:IntAsk-2}) can be re-arranged
to obtain
\begin{equation}
t=Lc_{0}+2c_{1}\:,\label{eq:IntT-2}
\end{equation}
with
\begin{eqnarray}
c_{0} & = & G(\mathbf{r})\textrm{ for all }\mathbf{r}\in\mathcal{A}_{o}\:,\label{eq:defc0-2}\\
c_{1} & = & \int_{0}^{\varsigma}g_{1}(z)dz\:.
\end{eqnarray}
\hfill{}$\blacksquare$

Finally we can broaden the field of application of Theorem 2 from
$\mathbb{P}^{d}$ to $\mathbb{P}_{\infty}^{d}\supset\mathbb{P}^{d}$.
We define $\mathcal{A}\in\mathbb{P}_{\infty}^{d}$ if $\mathcal{A}\subseteq\mathbb{R}^{d}$,
and $\mathcal{A}$ is the connected union of \textsl{countably many}
closed convex polytopes, and $\exists$ $\varepsilon\in\mathbb{R}_{>0}$
such that $\forall\lambda\in\left(0,\varepsilon\right)$ and for every
$\mathbf{r}\in\mathcal{A}$ the set $\textrm{int}\left[U\left(\mathbf{r},\lambda\right)\cap\mathcal{A}\right]$
is connected. In this case it is sufficient to consider the extension
of Eqs. (\ref{eq:IntE}) and (\ref{eq:IntE-1}) to the case of sums
over countable many faces. Note that even when $\mathcal{A}\in\mathbb{P}_{\infty}^{d}$
the theorem produce useful results only if $\mathfrak{L}(\mathcal{A})$
is positive definite.

\subsection{Application of both theorems to confined fluids\label{sub:Corolaries}}

Now we turn the attention to a fluid confined by a polytope $\mathcal{A}\in\mathbb{P}$
(the fluid was described in Sec. \ref{sec:PartitionFunc}) and analyze
the functions $\tau_{i}:\mathbb{P}\rightarrow\mathbb{R}$, $Z_{i}:\mathbb{P}\rightarrow\mathbb{R}$
and $\Xi_{M}:\mathbb{P}\rightarrow\mathbb{R}$. In particular, to
analyze $\tau_{i}$, one must first consider the Theorem 1 and thus
apply Theorem 2 to solve the integral in Eq. (\ref{eq:TauiE}). We
have obtained the following corollaries of the Theorems 1 and 2 that
apply to a confined fluid (relations are written for the case $d=3$
but other values of $d$ are easily included in our approach):
\begin{enumerate}
\item \label{enu:CorolTau}If a system of particles that interact via a
pair potential of finite range $\xi$ is confined in a polytope $\mathcal{A}\in\mathbb{P}^{3}$
such that $\mathfrak{L}(\mathcal{A})>2\varsigma$ with \foreignlanguage{english}{$\varsigma$}
taken from Eq. (\ref{eq:psi2}) for some positive integer $i>1$ then,
the $i$-th cluster integral $\tau_{i}:\mathbb{P}^{3}\rightarrow\mathbb{R}$
is a linear function in the variables $V$, $A$ and $\{L_{1},L_{2},...\}$
(the length of the edges of $\mathcal{A}$). It follows by replacing
in Eq. (\ref{eq:IntE}) the magnitudes $t$, $G$ with $\tau_{i}$,
$E_{1}$ and in Eqs. (\ref{eq:IntT}) to (\ref{eq:defc3}) the magnitudes
$t$, $G$, $c_{0}$, $c_{1}$, $c_{2,n}$ and $c_{3,n}$ with $\tau_{i}$,
$E_{1}$, $i!b_{i}$, $-i!a_{i}$, $i!c_{i,n}^{\textrm{e}}$ and $i!c_{i,n}^{\textrm{v}}$,
respectively. Then, $g(\mathbf{r})=E_{1}(\mathbf{r})-c_{0}$ (being
$c_{0}=E_{1}(\mathbf{r})$ with $\mathbf{r}\in\mathcal{A}_{o}$) and
the expression of the $i$-th cluster integral is
\begin{equation}
\tau_{i}/i!=V\, b_{i}-\, A\, a_{i}+\sum_{n\textrm{edges}}L_{n}c_{i,n}^{\textrm{e}}+\sum_{n\textrm{vertex}}c_{i,n}^{\textrm{v}}\;,\label{eq:TauiCorol}
\end{equation}
where the coefficients of $b_{i}$ and $a_{i}$ are independent of
the shape of $\mathcal{A}$ while $c_{i,n}^{\textrm{e}}$ is a function
of the dihedral angle in the $n$-th edge and $c_{i,n}^{\textrm{v}}$
is a function of the dihedral angles involved in the $n$-th vertex.
Besides, all the coefficients depends on the pair interaction potential,
the temperature and $i$. Finally, Eq. (\ref{eq:TauiZj}) shows that
$b_{1}=1$ and $a_{1}=c_{1,n}^{\textrm{e}}=c_{1,n}^{\textrm{v}}=0$.
\item \label{enu:CorolQ}If a system with $N$ particles that interact via
a pair potential of finite range $\xi$ is confined in a polytope
$\mathcal{A}\in\mathbb{P}^{3}$ such that $\mathfrak{L}(\mathcal{A})>2\varsigma$
with \foreignlanguage{english}{$\varsigma$} taken from Eq. (\ref{eq:psi2})
and $i$ replaced by $N$, then its canonical partition function is
polynomial on $V$, $A$ and $\{L_{1},L_{2},...\}$. It follows from
the previous corollary, Eqs. (\ref{eq:CPF}) and (\ref{eq:ZjTaui}).
From them the polynomials $Z_{N}$ and $Q$ can be explicitly obtained.
\item \label{enu:CorolGC}If a system with at most $M$ particles that interact
via a pair potential of finite range $\xi$ is confined in a polytope
$\mathcal{A}\in\mathbb{P}^{3}$ such that $\mathfrak{L}(\mathcal{A})>2\varsigma$
with \foreignlanguage{english}{$\varsigma$} taken from Eq. (\ref{eq:psi2})
and $i$ replaced by $M$ then, its (restricted) grand canonical partition
function is polynomial on $z$, $V$, $A$ and $\{L_{1},L_{2},...\}$.
This follows from the second corollary and Eqs. (\ref{eq:GPF})-(\ref{eq:GPF2}).
From them the polynomial $\Xi_{M}$ can be explicitly obtained.
\end{enumerate}
For practical purposes one introduce the mean value for the edge and
vertex coefficients $c_{i}^{\textrm{e}}=L_{\textrm{e}}^{-1}\sum_{n\textrm{edges}}L_{i}c_{i,n}^{\textrm{e}}$
and $c_{i}^{\textrm{v}}=N_{\textrm{v}}^{-1}\sum_{n\textrm{vertex}}c_{i,n}^{\textrm{v}}$,
respectively (with $L_{\textrm{e}}=\sum_{n\textrm{edges}}L_{n}$ and
$N_{\textrm{v}}$ the quantity of vertex in $\mathcal{A}$). Using
this mean values the Eq. (\ref{eq:Tauimean}) is re-written as 
\begin{equation}
\tau_{i}/i!=V\, b_{i}-A\, a_{i}+L_{\textrm{e}}c_{i}^{\textrm{e}}+N_{\textrm{v}}c_{i}^{\textrm{v}}\;.\label{eq:Tauimean}
\end{equation}
with $a_{1}=c_{1}^{\textrm{e}}=c_{1}^{\textrm{v}}=0$.

As a closing remark we can mention that for $d=1$ it is easy to demonstrate
that the conjectured relation (\ref{eq:psi3guess}) is true. Therefore,
one found that $\tau_{i}/i!=b_{i}L-2a_{i}$ apply to any $i<i_{max}$
with $i_{max}=\textrm{IntegerPart}\left(L/\varsigma\right)+1$ {[}see
Eq. (\ref{eq:IntE-2-1}){]}.

\subsection{Classification of $g$}

The three corollaries of Theorems 1 and 2 fix the structure of $\tau_{i}$,
$Z_{N}$ and $\Xi$, and provide explicit expressions that enables
to obtain the coefficients of $\tau_{i}$. To make a further step
in this direction we analyze the different scenarios for $E_{1}(\mathbf{r})$
and $g(\mathbf{r})$. At fixed $\mathbf{r}\in\mathcal{A}$ both functions
depend on the set $U(\mathbf{r},\varsigma)\cap\mathcal{A}$ throw
the shape of $U(\mathbf{r},\varsigma)\cap\partial\mathcal{A}$. Once
again, we mainly concentrate in the case $d=3.$ In terms of the partition
for $\mathcal{A}$ introduced in Eqs. (\ref{eq:APart}) and (\ref{eq:AskPart})
$\mathbf{r}\in\mathcal{A}$$\,\Rightarrow\,$$\mathbf{r}\in\mathcal{A}_{k}$
(for some $k=0,1,2,3$) and then $g(\mathbf{r})=g_{k}(\mathbf{r})$.
We have considered all the possible polyhedrons $\mathcal{A}$ with
$\mathfrak{L}(\mathcal{A})>2\varsigma$ and observed that there exist
countable many different cases for the geometry of $U(\mathbf{r},\varsigma)\cap\mathcal{A}$
and in each of this geometries $g(\mathbf{r})$ must be a smooth function
of $\mathbf{r}$ and the angles that describe the local shape of $\partial\mathcal{A}$.
In Table \ref{tab:Clasif} we present such a classification for $d=3$
and the case where at most three faces join in each vertex.
\begin{table}
\begin{centering}
\begin{tabular}{|c|c|c|c|c|c||c|c|c|c|c|c||c|c|c|c|c|c|}
\hline 
$k$ & $j$ & f & e & v &  & $k$ & $j$ & f & e & v &  & $k$ & $j$ & f & e & v & \tabularnewline
\hline 
\hline 
0 & 1 & 0 & 0 & 0 & \includegraphics[scale=0.5]{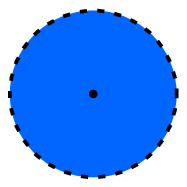} & 3 & 1 & 3 & 0 & 0 & \includegraphics[bb=0bp 0bp 70bp 70bp,scale=0.5]{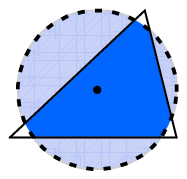} & 3 & 6 & 3 & 2 & 0 & \includegraphics[bb=0bp 0bp 70bp 70bp,scale=0.5]{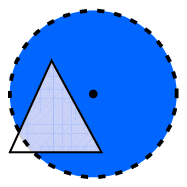}\tabularnewline
\hline 
1 & 1 & 1 & 0 & 0 & \includegraphics[bb=0bp 0bp 53bp 49bp,scale=0.5]{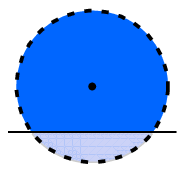} & 3 & 2 & 3 & 0 & 0 & \includegraphics[bb=0bp 0bp 70bp 70bp,scale=0.5]{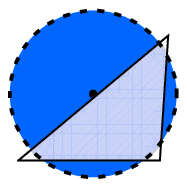} & 3 & 7 & 3 & 3 & 0 & \includegraphics[scale=0.5]{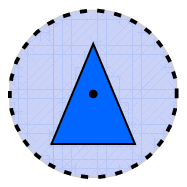}\tabularnewline
\hline 
2 & 1 & 2 & 0 & 0 & \includegraphics[bb=0bp 0bp 53bp 53bp,scale=0.5]{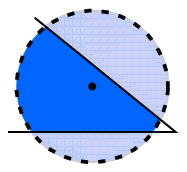} & 3 & 3 & 3 & 1 & 0 & \includegraphics[scale=0.5]{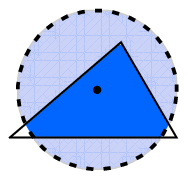} & 3 & 8 & 3 & 3 & 0 & \includegraphics[scale=0.5]{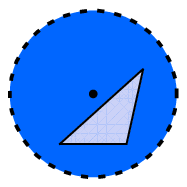}\tabularnewline
\hline 
2 & 2 & 2 & 0 & 0 & \includegraphics[scale=0.5]{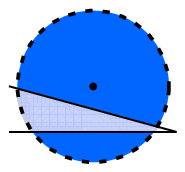} & 3 & 4 & 3 & 1 & 0 & \includegraphics[scale=0.5]{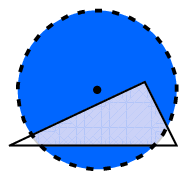} & 3 & 9 & 3 & 3 & 1 & \includegraphics[scale=0.5]{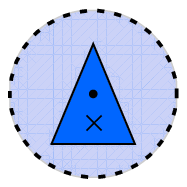}\tabularnewline
\hline 
2 & 3 & 2 & 1 & 0 & \includegraphics[scale=0.5]{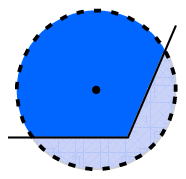} & 3 & 5 & 3 & 2 & 0 & \includegraphics[bb=0bp 0bp 53bp 51bp,scale=0.5]{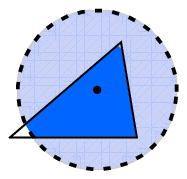} &  &  &  &  &  & \tabularnewline
\hline 
\end{tabular}
\par\end{centering}

\caption{Classification for the different shapes of $U(\mathbf{r},\varsigma)\cap\mathcal{A}$
(dark region of the draws, blue online). The $k$ index means that
$\mathbf{r}\in\mathcal{A}_{k}$ and $g(\mathbf{r})=g_{k}(\boldsymbol{r}^{(k)})$
with $g_{0}(\mathbf{r})=E_{1}(\mathbf{r})$, the $j$ index is introduced
to enumerate the different scenarios for the shape of $U(\mathbf{r},\varsigma)\cap\mathcal{A}$
at a given $k$. Columns f, e and v count the number of faces, edges
and vertex, respectively, that intersects the ball centered at $\mathbf{r}$
with radii $\varsigma$.\label{tab:Clasif}}
\end{table}
 There, the increasing complexity of the scenario with increasing
$k$ is apparent, there are only one case for $k=0,1$, three cases
for $k=2$ and nine cases for $k=3$. This behavior is a consequence
of the number of independent parameters involved, which are $0,1,3,5$
(none; $z$; $\boldsymbol{r}^{(2)}$ and $\beta$; and $\boldsymbol{r}^{(3)}$
and two -of the three- angles $\alpha$) for $k=0,1,2,3$, respectively.
This table shows e.g. that for a given $\tau_{i}$ we must know three
different expressions for $g_{2}=g_{2}(\boldsymbol{r}^{(2)},\alpha)$
if we expect to obtain the function $c_{i}^{\textrm{e}}(\alpha)$
which enable to describe all the possible $c_{i,n}^{\textrm{e}}$.
Finally, Table \ref{tab:Clasif} also apply for $d=1,2$. For $d=2$
we must restrict to $0\leq\textrm{f}\leq2$ and drop column $\textrm{v}$.
On the other hand, the columns $\textrm{f}$ and $\textrm{e}$ should
be renamed as: $\textrm{s}$ (sides) and $\textrm{v}$ (vertex), respectively.
For $d=1$ we must restrict to $0\leq\textrm{f}\leq1$, drop the columns
$\textrm{e}$ and $\textrm{v}$, and rename column $\textrm{f}$ as
$\textrm{v}$ (vertex).

\section{{\normalsize Extensions to non polytope shape of the confinement\label{sec:Extensions}}}

Once we have demonstrated the theorems and corollaries that reveal
the shape of $\tau(\mathcal{A})$ (and other magnitudes) for $\mathcal{A}\in\mathbb{P}$
with a characteristic length $\mathfrak{L}(\mathcal{A})>2\varsigma$,
we are ready to relax some of the constraints on the geometry of $\mathcal{A}$.
In a first step we analyze the case of curved faces. Focusing on $\mathbb{R}^{3}$,
if one of the planar faces is replaced with a curved surface with
constant and small curvature along it (cylindrical or spherical surfaces),
what can we expect to obtain for the behavior of $\tau(\mathcal{A})$?.
On the basis of the locality and translational invariance of $E_{1}(\mathbf{r})$,
the Eqs. (\ref{eq:TauiE}) and (\ref{eq:TauiCorol}), and the ideas
developed in the proof of Theorem 2 (case d=3), we expect that the
coefficients corresponding to the curved face, their vertex and edges
transform to functions of the principal radii of curvature of the
curved face $R_{I}$, $R_{II}$ (and the involved dihedral angles).
This functions of curvature radii could be expressed in powers of
its curvatures $R_{I}^{-1}$ and $R_{II}^{-1}$, being the independent
term of each series that corresponds to the planar case given in Eq.
(\ref{eq:TauiCorol}). For several faces with small constant curvature
one follow the same argument, in this case the edges (vertex) coefficients
will depend on the curvatures of the faces that meet in the edge (vertex)
and also on the involved dihedral angles. Related to this curvature
dependence, in the case of a hard spheres (HS) fluid the cluster integrals
$\tau_{2}$ and $\tau_{3}$ were obtained for a system confined by
spherical walls \cite{Urrutia_2008,Urrutia_2011_b,Urrutia_2011_be},
while $\tau_{2}$ was also evaluated for HS confined by cylindrical
and planar walls \cite{Urrutia_2008,Urrutia_2010b}. In addition,
the analysis of the HS fluid confined in the $d$-dimensional space
have provided explicit expressions of $\tau_{2}$ in the cases of
spherical and slit, confinement \cite{Urrutia_2010,Kim_2011}. Furthermore,
in the case of square-well particles $\tau_{2}$ was evaluated for
a spherical wall confinement \cite{Urrutia_2011}. In all these cases
the obtained expressions coincide with the above discussed general
picture.

A second generalization of the theorem corresponds to the periodic
boundary conditions which are frequently used in molecular dynamic
and montecarlo simulations. We consider a set of three non-coplanar
vectors $\left\{ \mathbf{v}_{1},\mathbf{v}_{2},\mathbf{v}_{3}\right\} $
that define the cell $C=\left\{ \mathbf{r}/\mathbf{r}=x_{1}\mathbf{v}_{1}+x_{2}\mathbf{v}_{2}+x_{3}\mathbf{v}_{3}\,,\,\forall\,0\leq x_{i}<1\right\} $,
a parallelepiped, and the corresponding vectors of the Bravais lattice
$\mathbf{v}=m_{1}\mathbf{v}_{1}+m_{2}\mathbf{v}_{2}+m_{3}\mathbf{v}_{3}$
with $m_{i}$ running over all the integers. The translation of the
cell over all the Bravais lattice vectors $\mathbf{v}$ tiles $\mathbb{R}^{3}$.
Note that systems with discrete translational symmetry in only two
or one or zero directions (the last case corresponds to the absence
of discrete translational symmetry) can be obtained in two ways; by
fixing $m_{i}=0$ in the definition of the Bravais lattice vectors
for one or two or three of the integer numbers $\left\{ m_{1},m_{2},m_{3}\right\} $
(which reduce the lattice dimension), or alternatively, by taking
the large cell limit in one or two or three of the lengths $\left\{ v_{1},v_{2},v_{3}\right\} $
with $v_{i}=\left|\mathbf{v}_{i}\right|$. We will take the second
point of view to analyze the periodic polyhedron problem in the framework
of the three dimensional Bravais lattice. Let us define the periodic
polyhedron $\mathcal{A}$ by giving, a set of lattice vectors $\left\{ \mathbf{v}_{1},\mathbf{v}_{2},\mathbf{v}_{3}\right\} $
that define $C$ and $\mathbf{v}$, a polyhedron $\mathcal{B}\subset C$
and the periodic array of polyhedrons formed by all the copies of
$\mathcal{B}$ over the lattice, $\mathcal{B}_{lat}$. Furthermore,
we assume that $\mathcal{B}_{lat}$ belongs to a kind of generalized
$\mathbb{P}$ space where the possibility of certain types of union
of infinite convex sets (that consistent with a Bravais lattice) are
allowed. We have strong evidence that shows that theorems 1 and 2
and their corollaries can be extended to include the case of a periodic
polytope $\mathcal{A}$. In this case the characteristic length $\mathfrak{L}\left(\mathcal{A}\right)$
should be redefined. On one side, one should consider the characteristic
length of the normal boundary of $\mathcal{A}$, $\mathfrak{L}'\left(\mathcal{A}\right)=\mathfrak{L}(\mathcal{B}_{lat})$,
given by Eq. (\ref{eq:LdeA}), that can be measured on a subset of
$\mathcal{B}_{lat}$ given by the cell and its $3^{3}-1$ neighbors.
On the other side, one should take into account the second characteristic
length of $\mathcal{A}$ which concerns the geometry of the periodic
cell, given by $\mathfrak{L}''\left(\mathcal{A}\right)=\min\left(v_{1},v_{2},v_{2}\right)$.
The minimum between both lengths $\mathfrak{L}'\left(\mathcal{A}\right)$
and $\mathfrak{L}''\left(\mathcal{A}\right)$ is the characteristic
length of the periodic polytope $\mathfrak{L}(\mathcal{A})$. Due
to at present, we have not a full demonstration of this generalization
to periodic polytopes, it must be considered a conjecture.

Other interesting extensions that should be analyzed in the future
are: multi-component systems, system with non-spherical interaction
potential, and external potentials that include other non-hard wall-particle
interactions.

\section{{\normalsize Thermodynamic properties\label{sec:Thermodynamics}}}

In Secs. \ref{sec:The-properties-of-Taui} and \ref{sec:polytope-theo}
it were derived new exact results on the statistical mechanics of
confined fluids inhomogeneously distributed in the space without any
particular spatial symmetry that applies both, to systems composed
of few particles, as well as to systems composed by many particles.
Recently, a thermodynamic approach that enables to study in a unified
way both types of systems was formulated \cite{Urrutia_2010b,Urrutia_2012}.
We apply here this thermodynamic approach to study the properties
of a fluid confined in a polytope in the framework of both, canonical
and grand canonical ensembles. This approach rest in two basic assumptions
made about the system: (1) it can be assigned to the system a free
energy, this can be done through the usual free-energy $\leftrightarrow$
partition-function relation which is a well established link between
statistical mechanics and thermodynamics, (2) under general ergodic
conditions there exist an identity between: the mean temporal value
of the properties of the system, those obtained from a thermodynamic
approach, and those found from mean ensemble value (the ensemble that
mimics the real constraints on the system must be considered). In
the rest of this section we will find and discuss the thermodynamic
relations, the equations of state (EOS), the relations between different
EOS, and their meaning.

For a system with a fixed number of particles, $N$, and temperature,
$T$, one must adopt the canonical ensemble. In this framework, the
connection between statistical mechanics and thermodynamics is
\begin{equation}
F=-\beta^{-1}\ln Q\:,\label{eq:F0}
\end{equation}
where $F$ is the Helmholtz free energy of the system. Besides, the
thermodynamic fundamental relations for $F$ that follows from the
first and second laws of thermodynamic through a Legendre transformation
are
\begin{equation}
F=U-TS\:,\label{eq:FhelmTh}
\end{equation}
\begin{equation}
dF=-S\, dT-dW_{\textrm{r}}\:.\label{eq:dFhelmTh}
\end{equation}
Here, $S=-\partial F/\partial T|_{N,\mathcal{A}}$ is the entropy
(the derivative is taken at fixed size and shape of the vessel), $dW_{\textrm{r}}=P_{W}dV$
is the total reversible work performed by the system on its environment
and $P_{W}$ is the pressure-for-work \cite{Urrutia_2010b}. For
a general vessel $\mathcal{A}$ with a complex shape an arbitrary
transformation may modify its size and shape. Thus, it is convenient
to fix a given transformation for $\mathcal{A}$ and its parametrization.
Using the $\lambda$ parameter one found $\mathcal{A}\rightarrow\mathcal{A}(\lambda)$,
$dV\rightarrow d\lambda\, dV/d\lambda$ and the first EOS of the system
\begin{equation}
P_{W}=-\left.\frac{\partial F}{\partial\lambda}\right|_{T,N}\left(\frac{dV}{d\lambda}\right)^{-1}\:.\label{eq:PwCan}
\end{equation}
It is interesting to note that once the expression for $F(T,\lambda)$
and $P_{W}$ are known an apparently thorough description of the thermodynamic
magnitudes involved in Eqs. (\ref{eq:FhelmTh}, \ref{eq:dFhelmTh})
is obtained. On the other hand, given that the system under study
is an inhomogeneous one $P_{W}$ is not the pressure in the fluid,
in fact, this simple observation shows that the obtained description
of the system in terms of $P_{W}$ alone is unsatisfactory because
it said almost nothing about the properties of the fluid itself.

To make further progress toward a more satisfactory thermodynamic
description of the system properties, we assume that $\mathcal{A}\in\mathbb{P}^{3}$
is a SB polyhedron and the confined fluid is such that the Corollaries
\ref{enu:CorolTau} and \ref{enu:CorolQ} in Sec.\ref{sub:Corolaries}
apply. Thus, based on Eq. (\ref{eq:TauiCorol}) we adopt the continuous
measures $X=\{V,A,\mathbf{L},\boldsymbol{\alpha}\}$ where bold symbols
$\mathbf{L}$ and $\boldsymbol{\alpha}$ are short notations for the
set of edges length and dihedral angles, $\alpha$. Naturally, these
measures can also be parametrized with $\lambda$ when it becomes
convenient. Besides, from Corollary \ref{enu:CorolQ} and Eqs. (\ref{eq:FhelmTh},
\ref{eq:dFhelmTh}), we obtain the explicit form of $F$, $S$, $U$
(from the statistical mechanical mean value recipe $U=\partial\beta F/\partial\beta|_{N,\mathcal{A}}$)
and $P_{W}$ in terms of $X$. We also found the following EOS 
\begin{equation}
P=-\left.\frac{\partial F}{\partial V}\right|_{T,N,X-V}\:,\quad\gamma=\left.\frac{\partial F}{\partial A}\right|_{T,N,X-A}\:,\label{eq:PyGamCan}
\end{equation}
\begin{equation}
\mathcal{T}_{n}=\left.\frac{\partial F}{\partial L_{n}}\right|_{T,N,X-L_{n}}\:,\quad\omega_{n}=\left.\frac{\partial F}{\partial\alpha_{n}}\right|_{T,N,X-\alpha_{n}}\:,\label{eq:TnysOmgCan}
\end{equation}
where the partial derivatives can be interpreted as the response of
$F$ to an small change of $V$ ($A$, $L_{n}$, $\alpha_{n}$) while
the other measures are kept constant. The Eqs. (\ref{eq:PyGamCan},
\ref{eq:TnysOmgCan}) may be seen simply as definitions for $P$,
$\gamma$, $\mathcal{T}_{n}$ and $\omega_{n}$. In this context,
each of these magnitudes is related to certain type of work, for example
$P$ is the work needed to change the volume of the system from $V$
to $V+dV$ (at constant $A,\mathbf{L}$ and $\boldsymbol{\alpha}$)
and $\omega_{n}$ is the work necessary to change the dihedral angle
$\alpha_{n}$ in $d\alpha_{n}$ (at all the measures $X$ but $\alpha_{n}$
constants, which affect both the edge and their vertex). However,
the Eqs. (\ref{eq:PyGamCan}, \ref{eq:TnysOmgCan}) also suggest the
meaning of $P$, $\gamma$ and $\mathcal{T}_{n}$; hence, $P$ should
be the pressure in the fluid, $\gamma$ its wall-fluid surface tension
and $\mathcal{T}_{n}$ the line tension corresponding to the $n$-th
edge. In favor of this interpretation one observe that when the fluid
develops a region with homogeneous density the pressure $P$ from
Eq. (\ref{eq:PyGamCan}) is the pressure of the fluid in this region.
This fact can be demonstrated using Theorems 1 and 2 and an approach
similar to that developed in Ref.\cite{Urrutia_2010b}. Thermodynamic
magnitudes $P$, $\gamma$, $\mathcal{T}_{n}$ and $\omega_{n}$ depend
on the adopted measures $X$ but are independent of the adopted transformation.
An equilibrium equation relates the different EOS,
\begin{equation}
P-P_{W}=q_{\gamma}\gamma+\left\langle q_{\mathcal{T}}\mathcal{T}\right\rangle +\left\langle q_{\omega}\omega\right\rangle \:,\label{eq:LplCan01}
\end{equation}
where the $q$ coefficients are of geometrical nature $q_{\gamma}=\frac{dA}{d\lambda}\left(\frac{dV}{d\lambda}\right)^{-1}$,
$\left\langle q_{\mathcal{T}}\mathcal{T}\right\rangle =\sum_{n\textrm{edges}}q_{\mathcal{T},n}\mathcal{T}_{n}$,
$\left\langle q_{\omega}\omega\right\rangle =\sum_{n\textrm{vertex}}q_{\omega,n}\omega_{n}$,
with $q_{\mathcal{T},n}=\left(\frac{dV}{d\lambda}\right)^{-1}\frac{dL_{n}}{d\lambda}$
and $q_{\omega,n}=\left(\frac{dV}{d\lambda}\right)^{-1}\frac{d\alpha_{n}}{d\lambda}$.
The Eq. (\ref{eq:LplCan01}) is very similar to the Laplace Equation
for a drop that express the equilibrium between the inner pressure
(in the liquid phase), the outer pressure (in the vapor phase) and
the liquid-vapor surface tension. Hence, we say that Eq. (\ref{eq:LplCan01})
is a Laplace-like relation. For the case of $\frac{dV}{d\lambda}=0$
where Eqs. (\ref{eq:PwCan}) and (\ref{eq:LplCan01}) have none sense
one may define $dW=-\gamma_{W}dA$, $\gamma_{W}=\left.\frac{\partial F}{\partial\lambda}\right|_{T}\left(\frac{dA}{d\lambda}\right)^{-1}$
and the Laplace-like relation gives 
\begin{equation}
\gamma_{W}-\gamma=\left\langle r_{\mathcal{T}}\mathcal{T}\right\rangle +\left\langle r_{\omega}\omega\right\rangle \:,\label{eq:LplCanGam}
\end{equation}
where $\left\langle r_{\mathcal{T}}\mathcal{T}\right\rangle =\sum_{n\textrm{edges}}r_{\mathcal{T},n}\mathcal{T}_{n}$
$\left\langle r_{\omega}\omega\right\rangle =\sum_{n\textrm{vertex}}r_{\omega,n}\omega_{n}$,
with $r_{\mathcal{T},n}=\left(\frac{dA}{d\lambda}\right)^{-1}\frac{dL_{n}}{d\lambda}$
and $r_{\omega,n}=\left(\frac{dA}{d\lambda}\right)^{-1}\frac{d\alpha_{n}}{d\lambda}$.
Eq. (\ref{eq:LplCanGam}) makes explicit that the usual approach to
the surface tension which consist in measure the reversible work of
doing a transformation at constant volume does not give the thermodynamic
surface tension $\gamma$, even it measure $\gamma_{W}$ for this
adopted transformation. Now we focus on the case of some simpler shapes
for $\mathcal{A}$ to obtain more explicit expressions of Eq. (\ref{eq:LplCan01}).
For a convex polyhedron $\mathcal{A}$ which is also regular all its
edges and vertex are equivalent (equal dihedral angles and edges length),
one can make a re-scaling transformation (at fixed angles) by multiplying
the vectors that fix the position of each vertex by a parameter $\kappa\approx1$.
In this case, for an initial state $X=\left\{ V_{o},A_{o},\mathbf{L}_{o}\right\} $
we found $V'=3\kappa^{2}V_{o}$, $A'=2\kappa A_{o}$, $L_{n}'=L_{n,o}$,
$q_{\gamma}=\frac{2A_{o}}{3\kappa V_{o}}$, $\left\langle q_{\mathcal{T}}\mathcal{T}\right\rangle =\mathcal{T}\frac{L_{\textrm{e},o}}{3\kappa^{2}V_{o}}$
(with $\mathcal{T}=\mathcal{T}_{n}$ for any $n$-edge) and $\left\langle q_{\omega}\omega\right\rangle =0$.
On the other hand, for the case of polyhedron $\mathcal{A}$ such
that its complement $\mathcal{A}^{c}$ is convex and regular (if $\mathcal{A}^{c}$
is convex we say that $\mathcal{A}$ is com-convex), starting with
an initial state $X=\{V=V_{\infty}-V_{o},A_{o},\mathbf{L}_{o}\}$,
we obtain $V'=-3\kappa^{2}V_{o}$, $A'=2\kappa A_{o}$, $L_{n}'=L_{n,o}$,
which produce a minus sign in $q_{\gamma}=-\frac{2A_{o}}{3\kappa V_{o}}$
and $\left\langle q_{\mathcal{T}}\mathcal{T}\right\rangle =-\mathcal{T}\frac{L_{\textrm{e},o}}{3\kappa^{2}V_{o}}$.
By fixing $\kappa=1$ the Eq. (\ref{eq:LplCan01}) for both, the convex
and com-convex regular cases, reduce to
\begin{equation}
\left(P-P_{W}\right)\times\textrm{sg}=\gamma\,\frac{2A_{o}}{3V_{o}}+\mathcal{T}\,\frac{L_{\textrm{e},o}}{3V_{o}}\:,\label{eq:LplCan02}
\end{equation}
where $\textrm{sg}$ gives the sign of $\frac{dV(k)}{dk}$ (which
is $1$ or $-1$ for the convex and com-convex cases, respectively).
This equation applies not only to a cuboidal confinement, both for
the fluid in the cuboid and for the fluid surrounding it, but also,
for the tetrahedron and other platonic solids (octahedron, dodecahedron
and icosahedron), evenmore it also applies to the cuboctahedron and
the icosidodecahedron confinements (see %
\footnote{http://en.wikipedia.org/wiki/Platonic\_solid%
} for geometrical details). It is interesting to note that for the
above analyzed cases, where Eq. (\ref{eq:LplCan02}) apply, $P_{W}$
reduce to the mean pressure at the wall (wall-pressure) related with
the mean density on the wall $\rho_{\textrm{w}}$ through the contact
theorem 
\begin{equation}
P_{W}=\beta^{-1}\rho_{\textrm{w}}\:.\label{eq:contactTh}
\end{equation}

Open systems with inasmuch $M$ particles and $T$, $\mu$ -or $z$-
and $\mathcal{A}$ fixed must be analyzed in the framework of the
grand canonical ensemble. In this ensemble the connection between
statistical mechanics and thermodynamics is
\begin{equation}
\Omega=-\beta^{-1}\ln\Xi\:,\label{eq:Omega0}
\end{equation}
while the thermodynamic fundamental relations that follow from the
first and second laws are
\begin{equation}
\Omega=U-TS-\mu N\:,\label{eq:OmegaTh}
\end{equation}
\begin{equation}
d\Omega=-S\, dT-dW_{\textrm{r}}-N\, d\mu\:.\label{eq:dOmegaTh}
\end{equation}
By introducing a parametrization for the cavity transformation one
find
\begin{equation}
P_{W}=-\left.\frac{\partial\Omega}{\partial\lambda}\right|_{T,z}\left(\frac{dV}{d\lambda}\right)^{-1}\:.\label{eq:PwGCan}
\end{equation}
For the case of $\mathcal{A}\in\mathbb{P}^{3}$ a SB polyhedron and
a fluid such that the Corollaries \ref{enu:CorolTau} to \ref{enu:CorolGC}
apply, we use the same measures $X$ adopted above. Besides, from
Corollary \ref{enu:CorolGC} and using Eqs. (\ref{eq:GPF}) to (\ref{eq:Taui})
we obtain the explicit form of $\Omega$ and $P_{W}$ in terms of
$X$. The expressions for $S$, $U$ and the EOS in this GCE are easily
found by replacing $F$ with $\Omega$ in Eqs. (\ref{eq:PwCan}) to
(\ref{eq:TnysOmgCan}). For the other EOS we obtain
\begin{equation}
P=-\left.\frac{\partial\Omega}{\partial V}\right|_{T,z,X-V}\:,\quad\gamma=\left.\frac{\partial\Omega}{\partial A}\right|_{T,z,X-A}\:,\label{eq:PyGamGCan}
\end{equation}
\begin{equation}
\mathcal{T}_{n}=\left.\frac{\partial\Omega}{\partial L_{n}}\right|_{T,z,X-L_{n}}\:,\quad\omega_{n}=\left.\frac{\partial\Omega}{\partial\alpha_{n}}\right|_{T,z,X-\alpha_{n}}\:.\label{eq:TnysOmgGCan}
\end{equation}
As can be seen, several results are similar to that found for the
canonical ensemble case and thus we only draw about some features
of this GCE study. For example, from the EOS one obtain the equilibrium
relation
\begin{equation}
P-P_{W}=q_{\gamma}\gamma+\left\langle q_{\mathcal{T}}\mathcal{T}\right\rangle +\left\langle q_{\omega}\omega\right\rangle \:,\label{eq:LplGCan01}
\end{equation}
this Laplace-like relation is the same that Eq. (\ref{eq:LplCan01})
and again $P$ is the pressure in the fluid, which coincides with
the pressure in a region with homogeneous density {[}see details in
Eq. (\ref{eq:LplCan01}) or Eqs. (\ref{eq:LplCan02}, \ref{eq:contactTh})
for the specific case of regular polygons{]}. Other usual thermodynamic
and statistical mechanical relations that are valid even for the case
of inhomogeneous systems with a finite $M$ value are
\begin{equation}
N\equiv\left\langle N\right\rangle =-z\left.\frac{\partial\,\beta\Omega}{\partial z}\right|_{T,\mathcal{A}}\:,\label{eq:N}
\end{equation}
\begin{equation}
\sigma_{_{N}}\!^{2}\equiv\left\langle N^{2}\right\rangle -N^{2}=z\left.\frac{\partial N}{\partial z}\right|_{T,\mathcal{A}}\:,\label{eq:SgNsqr}
\end{equation}
where $\sigma_{_{N}}$, the standard deviation in the mean number
of particles $N$, quantifies its spontaneous fluctuation. Note that
derivatives with respect to $z$ are taken with the fluid in a fixed
region $\mathcal{A}$ (all the $X$ measures fixed).

\subsection{The low $z$ expansion}

To study the thermodynamic behavior of the confined open fluid in
the low $z$ regime one express Eq. (\ref{eq:Omega0}) as a power
series in $z$ {[}see also Eq. (\ref{eq:GPF2}){]}. We introduce the
vector of coefficients $\mathbf{b}_{i}=(b_{i},a_{i},c_{i,1}^{\textrm{e}},\cdots,c_{i,1}^{\textrm{v}},\cdots)$
which depends on $\boldsymbol{\alpha}$, and the vector of extensive-like
magnitudes $Y=\{V,A,\mathbf{L},\mathbf{1}_{\textrm{v}}\}$ with $\mathbf{1}_{\textrm{v}}$
the vector of $N_{\textrm{v}}$ components all of them equal to one
being $N_{\textrm{v}}$ the number of vertex. Hence
\begin{equation}
\frac{\tau_{i}}{i!}=Y\cdot\mathbf{b}_{i}\:,\label{eq:Tauvec}
\end{equation}
an expression equivalent to Eqs. (\ref{eq:TauiCorol}) and (\ref{eq:Tauimean}).
Using Eq. (\ref{eq:TauiZj}) we can write 

\begin{equation}
\beta\Omega=-\sum_{i\geq1}\frac{\tau_{i}}{i!}z^{i}=-\sum_{i=1}^{M}z^{i}\, Y\cdot\mathbf{b}_{i}+O_{M+1}\left(z\right)\:,\label{eq:OmegaSrz}
\end{equation}
which is exact to order $M$. Notably, Eq. (\ref{eq:OmegaSrz}) shows
that to this order $\Omega$ is linear in the extensive-like magnitudes
in $Y$, on the other hand, Eq. (\ref{eq:Tauimean}) shows that to
this order $\Omega$ is also linear in $V$, $A$, $L_{\textrm{e}}$
and $N_{\textrm{v}}$. To the same order we found

\begin{equation}
N\doteq\sum_{i=1}^{M}iz^{i}\, Y\cdot\mathbf{b}_{i}\:,\label{eq:NnSrz}
\end{equation}
\begin{equation}
\sigma_{_{N}}\!^{2}\doteq\sum_{i=1}^{M}i^{2}z^{i}\, Y\cdot\mathbf{b}_{i}\:.\label{eq:SgmSrz}
\end{equation}
Therefore, in Eqs. (\ref{eq:OmegaSrz})-(\ref{eq:SgmSrz}) and to
order $M$ one can separate each component. For example the Eq. (\ref{eq:OmegaSrz})
is
\begin{equation}
\Omega\doteq\Omega_{b}+\Omega_{s}+\Omega_{l}+\Omega_{p}\:,\label{eq:OmegaComp}
\end{equation}
with the volumetric part of the grand-potential $\Omega_{b}$, and
the grand potential of surface, line and points $\Omega_{s}$, $\Omega_{l}$
and $\Omega_{p}$, respectively. Each of they correspond to the sum
of terms proportional to $V$, $A$, $L_{\textrm{e}}$, $N_{\textrm{v}}$;
and is trivially related with a grand potential density (per unit
volume, area, etc.) 
\begin{equation}
P_{b}=-\frac{\Omega_{b}}{V}=-\left.\frac{\partial\Omega}{\partial V}\right|_{z,T,X-V}\doteq\beta^{-1}\sum_{i=1}^{M}z^{i}b_{i}\:,\label{eq:Pb}
\end{equation}
\begin{equation}
\gamma_{\infty}=\frac{\Omega_{s}}{A}=\left.\frac{\partial\Omega}{\partial A}\right|_{z,T,X-A}\doteq\beta^{-1}\sum_{i=2}^{M}z^{i}a_{i}\:,\label{eq:GamInf}
\end{equation}
\begin{equation}
\mathcal{T}_{\infty}=\frac{\Omega_{l}}{L_{\textrm{e}}}\doteq-\beta^{-1}\sum_{i=2}^{M}z^{i}c_{i}^{\textrm{e}}\:,\label{eq:TauInf}
\end{equation}
\begin{equation}
\nu=\frac{\Omega_{p}}{N_{\textrm{v}}}\doteq-\beta^{-1}\sum_{i=2}^{M}z^{i}c_{i}^{\textrm{v}}\:.\label{eq:Nu}
\end{equation}
Note that the right hand side equality in Eqs. (\ref{eq:Pb}) to
(\ref{eq:Nu}) is up to order $M$ in $z$. Here, Eqs. (\ref{eq:Pb})
and (\ref{eq:GamInf}) coincide with work terms definitions (\ref{eq:PyGamGCan})
while the Eqs. (\ref{eq:TauInf}) and (\ref{eq:Nu}) may be understood
as mean works. From the right hand side term of Eq. (\ref{eq:Pb})
one find that $P_{b}$ is the pressure of the bulk system (with the
same $T$ and $z$). This series was studied by Mayer and others (see
Ref. \cite{McMillan_1945} and a more complete survey in \cite[p.122]{Hill1956})
in their approach to the virial equation of state for the bulk system.
On the other hand, from the right hand side term in Eq. (\ref{eq:GamInf})
one find that $\gamma_{\infty}$ is the (wall-fluid) surface tension
for the bulk fluid in contact with an infinite planar wall. This series
is consistent with that obtained by Sokolowski and Stecki \cite{Sokolowski_1979},
a question that will be briefly discussed at the end of this section.
Furthermore, $P_{b}$ and $\gamma_{\infty}$ clearly corresponds to
$P$ and $\gamma$ from Eq. (\ref{eq:PyGamGCan}) but also to the
CE Eqs. (\ref{eq:PyGamCan}), which validates our thermodynamical
approach. The additional work terms, to order $M$ in $z$ are {[}from
Eq. (\ref{eq:TnysOmgGCan}){]}
\begin{equation}
\mathcal{T}_{n}\doteq-\beta^{-1}\sum_{i=2}^{M}z^{i}c_{i,n}^{\textrm{e}}\:,\label{eq:Taun}
\end{equation}
\begin{equation}
\omega_{n}\doteq-\beta^{-1}\sum_{i=2}^{M}z^{i}\left(L_{n}\left.\frac{\partial c_{i,n}^{\textrm{e}}}{\partial\alpha_{n}}\right|+\left.\frac{\partial c_{i,p}^{\textrm{v}}}{\partial\alpha_{n}}\right|+\left.\frac{\partial c_{i,q}^{\textrm{v}}}{\partial\alpha_{n}}\right|\right)_{z,T,X-\alpha_{n}},\label{eq:smallomega}
\end{equation}
where $p$ and $q$ are the vertex at the endpoints of the $n$-th
edge. Eq. (\ref{eq:Taun}) corresponds to the line tension on the
$n$-th edge. Thus, one can prove {[}using Eq. (\ref{eq:Tauimean}){]}
that $\mathcal{T}_{\infty}=\left\langle \mathcal{T}_{n}\right\rangle =L_{\textrm{e}}^{-1}\sum_{n\textrm{edges}}\mathcal{T}_{n}L_{n}$,
i.e., $\mathcal{T}_{\infty}$ is a mean-work term (the work needed
to increase the edges length in $dL_{\textrm{e}}$ of an edge with
mean properties) it is the mean line tension. On the other hand, 
\begin{equation}
\omega_{n}=L_{n}\left.\frac{\partial\mathcal{T}_{n}}{\partial\alpha_{n}}\right|_{z,T,X-\alpha_{n}}+\left.\frac{\partial(N_{\textrm{v}}\nu)}{\partial\alpha_{n}}\right|_{z,T,X-\alpha_{n}}\:.\label{eq:wn}
\end{equation}
In terms of the densities the Eq. (\ref{eq:OmegaComp}) transforms
to
\begin{equation}
\Omega\doteq-VP_{b}+A\gamma_{\infty}+L_{\textrm{e}}\mathcal{T}_{\infty}+N_{\textrm{v}}\nu\:,\label{eq:Omgsepdens}
\end{equation}
which is a demonstration that the grand potential is a homogeneous
function of the so called extensive variables $V$, $A$, $L_{\textrm{e}}$
and $N_{\textrm{v}}$ (at least to order $M$ in powers of $z$).
Note that this is a central assumption in the thermodynamic and statistical
mechanics theories of macroscopic systems composed by many particles
and spatially distributed following strong symmetries. On the opposite,
here it is demonstrated for an inhomogeneous system, may be composed
by few particles, without particular translational or rotational spatial
symmetries. Naturally, explicit expression for several excess grand
potential can also be obtained, for example the (over-bulk) surface
excess grand-potential density is $\bar{\gamma}\doteq(\Omega+VP_{b})/A=\beta^{-1}\sum_{i=2}^{M}z^{i}\left(a_{i}-c_{i}^{\textrm{e}}L_{\textrm{e}}/A-c_{i}^{\textrm{v}}N_{\textrm{v}}/A\right)$
which is a rough estimate of the surface tension. Introducing the
$\lambda$ parametrization for a general vessel transformation we
found the Laplace-like equilibrium relation
\begin{equation}
P_{b}-P_{W}=q_{\gamma}\gamma_{\infty}+\left\langle q_{\mathcal{T}}\mathcal{T}\right\rangle +\left\langle q_{\omega}\omega\right\rangle \:,\label{eq:Lpl01GC}
\end{equation}
equivalent to Eq. (\ref{eq:LplGCan01}), but in this case we found
the expression and meaning of each term to order $M$ in $z$. Again,
for the particular cases analyzed below Eq. (\ref{eq:LplCanGam})
we obtain
\begin{equation}
\left(P_{b}-P_{W}\right)\times\textrm{sg}=\gamma_{\infty}\,\frac{2A}{3V_{o}}+\mathcal{T}_{\infty}\,\frac{L_{\textrm{e}}}{3V_{o}}\:,\label{eq:Lpl02GC}
\end{equation}
where $\mathcal{T}_{\infty}=\mathcal{T}_{n}=\mathcal{T}$ is a function
of the dihedral angle {[}details are given below Eq. (\ref{eq:LplCan02}){]}.
This Eq. suggest the following procedure to obtain information about
$\gamma_{\infty}$ and $\mathcal{T}_{\infty}$. Consider $\tilde{\gamma}$
defined below
\begin{equation}
\tilde{\gamma}=\frac{3V_{o}}{2A}\textrm{sg}\times\left(P_{b}-P_{W}\right)=\gamma_{\infty}+\mathcal{T}_{\infty}\,\frac{L_{\textrm{e}}}{2A}\:,\label{eq:fitting}
\end{equation}
all the magnitudes between the equal signs are simple to measure in
a molecular dynamic simulation of the open system. In particular Eq.
(\ref{eq:contactTh}), which remains valid in the GCE, provides a
simple way to evaluate $P_{W}$. Hence, one can fix the vessel shape,
$T$ and a small $z$ value, and then do measures of $\tilde{\gamma}$
along simulations for several different sizes of the vessel. Plotting
$\tilde{\gamma}$ vs. $\frac{L_{\textrm{e}}}{2A}$ and fitting the
points with a linear regression we obtain from the ordinate and the
slope $\gamma_{\infty}$ and $\mathcal{T}_{\infty}$, respectively.
If we repeat the procedure for different values of $z$ we obtain
a table of values for $\gamma_{\infty}(z)$ and $\mathcal{T}_{\infty}(z)$
which enable to evaluate the coefficients $a_{i}$ and $c_{i}^{\textrm{e}}$
through a second fit in this case with a polynomial function.

The splitting of $\Omega$ in terms of its components to order $M$
in $z$, given in Eqs. (\ref{eq:OmegaComp}) and (\ref{eq:Omgsepdens}),
can also be done for Eq. (\ref{eq:NnSrz}). It gives 
\begin{equation}
\left(\rho-\rho_{b}\right)V=A\bar{\Gamma}=A\Gamma+L_{\textrm{e}}\Gamma_{\textrm{e}}+N_{\textrm{v}}\Gamma_{\textrm{v}}\doteq\sum_{i=2}^{M}iz^{i}\left(-a_{i}A+L_{\textrm{e}}c_{i}^{\textrm{e}}+N_{\textrm{v}}c_{i}^{\textrm{v}}\right)\:,\label{eq:Adsorp}
\end{equation}
where $\rho=N/V$ is the mean number density, $\rho_{b}\doteq\sum_{i=1}^{M}iz^{i}b_{i}$
is bulk density, $\bar{\Gamma}$ is the total effective adsorption,
while the area, edge and vertex adsorptions are $\Gamma$, $\Gamma_{\textrm{e}}$
and $\Gamma_{\textrm{v}}$, respectively. On the other hand, the same
procedure applied to the fluctuation in Eq. (\ref{eq:SgmSrz}) produce
\begin{equation}
\sigma_{_{N}}\!^{2}-\sigma_{b}\!^{2}=A\bar{s}=As+L_{\textrm{e}}s_{\textrm{e}}+N_{\textrm{v}}s_{\textrm{v}}\doteq\sum_{i=2}^{M}i^{2}z^{i}\left(-a_{i}A+L_{\textrm{e}}c_{i}^{\textrm{e}}+N_{\textrm{v}}c_{i}^{\textrm{v}}\right)\:,\label{eq:Fluct}
\end{equation}
with $\sigma_{b}\!^{2}\doteq V\sum_{i=1}^{M}i^{2}z^{i}b_{i}$ the
fluctuation in the number of particles in the bulk, $\bar{s}$ an
excess effective fluctuation per unit area while $s$, $s_{\textrm{e}}$,
and $s_{\textrm{v}}$ are the components of surface-area, edges length
and vertex of the number density fluctuation. Particular attention
deserve such polytopes for which $ $$\mathfrak{L}(\mathcal{A})$
is infinite. They are the unbounded polytopes with one face, or one
edge, or one vertex. In this cases $M$ can take any positive integer
value and thus it be made as larger as one wishes. Therefore all the
series in powers of $z$ given between Eq. (\ref{eq:OmegaSrz}) and
Eq. (\ref{eq:Fluct}) are valid to any order. Besides, the linear
decomposition of several functions in its basic extensive measures
$Y\left(\mathcal{A}\right)$ is also exact to any order. In particular,
the grand potential becomes an homogeneous function of the extensive
measures {[}see Eq. (\ref{eq:Omgsepdens}){]}.

The structure of Eqs. (\ref{eq:OmegaComp}-\ref{eq:Nu}, \ref{eq:Omgsepdens},
\ref{eq:Adsorp}) and (\ref{eq:Fluct}) show that the relations between
the functions 
\begin{equation}
\left(\beta P_{b}-z,\,\rho_{b}-z,\,\sigma_{b}\!^{2}-z\right)\:\textrm{ and }b_{i>1},\label{eq:funcset01}
\end{equation}
 (where $\beta P_{b}-z$, $\rho_{b}-z$ and $\sigma_{b}\!^{2}-z$
are excess magnitudes with respect to the ideal gas system) is the
same that the relation between the functions
\begin{equation}
\left(\beta\bar{\gamma},\,-\bar{\Gamma},\,-\bar{s}\right)\:\textrm{ and }a_{i}-c_{i}^{\textrm{e}}L_{\textrm{e}}A^{-1}-c_{i}^{\textrm{v}}N_{\textrm{v}}A^{-1},\label{eq:funcset02}
\end{equation}
while the same apply to these other sets of functions and coefficients
\begin{equation}
\left(\beta\gamma_{\infty},\,-\Gamma,\,-s\right)\:\textrm{ and }a_{i},\label{eq:funcset03}
\end{equation}
\begin{equation}
\left(-\beta\mathcal{T}_{\infty},\,\Gamma_{\textrm{e}},\, s_{\textrm{e}}\right)\:\textrm{ and }c_{i}^{\textrm{e}},\label{eq:funcset04}
\end{equation}
\begin{equation}
\left(-\beta\nu,\,\Gamma_{\textrm{v}},\, s_{\textrm{v}}\right)\:\textrm{ and }c_{i}^{\textrm{v}}.\label{eq:funcset05}
\end{equation}
These simple observations show for example that $-\beta\gamma_{\infty}(\Gamma)$
and the dependence of $\beta P_{b}-z$ on $\rho_{b}-z$ is the same
except that $-a_{i}$ in the first case must be replaced by $b_{i}$
in the second case. 

Trivial series manipulation (inversion and composition) enables to
obtain explicit expressions for the power series (to order $M$) of
$P_{b}(\rho)$, $P_{b}(\rho_{b})$, $\gamma_{\infty}(\rho_{b})$,
$\bar{\gamma}(\rho)$, $\gamma(\Gamma)$, $\gamma(s_{\textrm{e}})$,
$\Gamma(\rho_{b})$, $\mathcal{T}_{\infty}(\rho)$, $\mathcal{T}_{\infty}(\rho_{b})$
and others. Several examples of these series discussed in terms of
the formal symmetries given in Eqs. (\ref{eq:funcset01}) to (\ref{eq:funcset05})
are explicitly evaluated in Appendix \ref{sec:AppendixB}. For fixed
$M$ we have verified that the series for $P_{b}(\rho_{b})$ (to order
$M$) is the usual virial series (see for example Ref. \cite{Hill1956}),
while $\gamma_{\infty}(\rho_{b})$ and $\Gamma(\rho_{b})$ reproduce
the approach of Bellemans \cite{Bellemans_1962} and subsequent improvements
of Sokolowski and Stecki \cite{Sokolowski_1979} which have developed
a complete theory of virial expansion for surface thermodynamic properties
for the case of a fluid in contact with a planar infinite wall. As
far as we known the series representation of any other of the discussed
functional relations were never analyzed before.

\section{{\normalsize Final Remarks\label{sec:Conclu}}}

The obtained polynomial structure of the CE and GCE partition functions
of few- and many-body inhomogeneous fluid-like systems confined by
polytopes is interesting in several ways. It shows that the statistical
mechanical properties of many of these fluid-like systems with a fixed
number of particles can be exactly described if the involved coefficients
(which depend on $T$) are known. This consideration also applies
to open systems if we conveniently restrict the maximum number of
particles considered. The key point for these results was the analysis
of the geometrical properties of the reducible cluster integrals for
inhomogeneous systems. It is interesting to mention that from the
decade of 1950 until present this integrals (introduced by Mayer and
Mayer \cite{Mayer1940} for the study of homogeneous fluids) were
not the focus of much interest and were almost ignored in the study
of inhomogeneous fluids. On the other hand, given that our approach
to Theorems 1, 2 and their corollaries is constructive we found explicit
integral expressions for the coefficients. In a work in progress
we evaluate some of the unknown lower order coefficients for a system
of hard sphere particles.

The results for the partition functions were complemented with a statistical
mechanical and thermodynamical recipe that, based on the linear decomposition
of cluster integrals enable to find the exact properties of the confined
inhomogeneous fluid (for both the open and closed systems). This fact
is notable because the studied confined inhomogeneous systems, which
are far away from the many particles in large volumes condition, have
not any particular spatial symmetry and thus the very basic assumptions
of the usual approach to thermodynamics theory are not satisfied.
Furthermore, it was shown that in the low density (and low $z$) regime
the obtained grand potential is a homogeneous function of the extensive
variables, which is a consequence of the linear decomposition of the
cluster integrals in their extensive components (for $d=3$ they are
volume, boundary surface area, edges length and vertex number). Even
more, it was found a generalized version of virial series EOS, that
provides expressions for pressures, wall-fluid surface tension, line
tension, adsorption, etc. All this findings strongly suggest that
it is possible to build a wider scope formulation of both, statistical
mechanics and thermodynamics, theories. These generalized formulations
should concern to systems with or without symmetries, involving from
few- to many-bodies, and constrained to regions of any size. 

\begin{figure}
\centering{}\includegraphics[width=6cm]{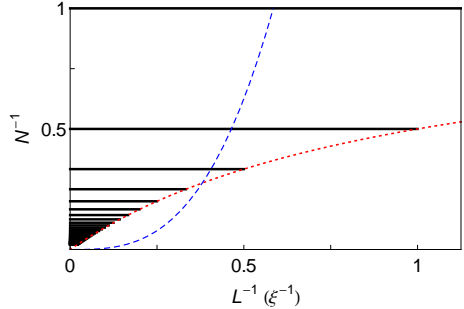}\caption{The size vs. number domain for the system of $N$ particles in a cube
with side $L$. Below the dotted curve the Corollaries 2 and 3 do
not apply and thus the exact structure of the partition functions
is unknown.\label{fig:Size}}
\end{figure}
In Fig. \ref{fig:Size} it is represented the domain where the CE
(GCE) partition function is a polynomial for the case of $N$ particles
(at most $N$ particles) confined in a cube with side $L$. Broad
straight lines show this domain for $Z_{N}$ and $\Xi_{N}$ for different
(fixed) number of particles. Each line extends from the infinite dilution
limit $L^{-1}\rightarrow0$ to the maximum density for which the Corollaries
given in Sec. \ref{sub:Corolaries} apply. This maximum density condition
is given by $N^{-1}=(1+L/\xi)^{-1}$ (in accord with Eq. (\ref{eq:psi3guess})
and Sec. \ref{sub:Corolaries}) and is drawn in dotted line. The large
systems limit corresponds to $V\rightarrow\infty$, $N\rightarrow\infty$
and coincides with the origin. However, to analyze the thermodynamic
limit one must reach large systems by following a constant density
path, which correspond to a dependence $N^{-1}\propto L^{-3}$. In
the figure it is plotted for a small value $\rho=0.2\,\xi^{-3}$ with
a dashed line. It is clear that our results about CE are insufficient
to analyze the thermodynamic limit because for large volumes (small
abscissa values) the highest density under which the structure of
the partition function is known gives a linear relation $N^{-1}\simeq L^{-1}$.
On the basis of this result is the nature of Theorems 1 and 2 which
show that the end of validity of Eq. (\ref{eq:TauiCorol}) relates
with the existence of a cluster configuration that is capable to percolate
the cavity $\mathcal{A}$.

An attempt to analyze the polytopes $\mathcal{A}$ such that $2\varsigma\gtrsim\mathfrak{L}(\mathcal{A})$
shows two folds. On one hand, that for some of these polytopes the
Eq. (\ref{eq:TauiCorol}) remains valid (as well as, apply the conclusions
of Corollaries 2 and 3). In this case, appears a less restrictive
condition $2\varsigma'<\mathfrak{L}(\mathcal{A})$ with $\varsigma'<\varsigma$
that replace $2\varsigma<\mathfrak{L}(\mathcal{A})$. On the other
hand, that the end of applicability of Eq. (\ref{eq:TauiCorol}) relates
with the non-universal behavior of $\tau_{N}$ due to the existence
of a non-analytic term {[}which is identically null for $2\varsigma<\mathfrak{L}(\mathcal{A})${]}
that depends on the shape of $\mathcal{A}$ in a more complex way.
The study of this term for some simple confinements is interesting
because it may enlighten how to describe the system properties in
the thermodynamic limit.

The core of the formal result presented in this work relies on Theorems
1 and 2. As was mentioned, Theorem 2 for $d=1,2,3$ and the conjectured
generalization to $\mathbb{R}^{d}$ given in Eq. (\ref{eq:IntEd})
strongly resemble well known results of integral geometry. In particular,
the expressions given in Eqs. (\ref{eq:IntEd}, \ref{eq:IntE}, \ref{eq:IntE-1})
and (\ref{eq:IntE-2-1}) look like the combination of Hadwiger's characterization
theorem and The general kinematic formula (Theorems 9.1.1 and 10.3.1
in pp. 118 and 153 of \cite{Klain1991}, respectively) as it were
applied to a polytope and a ball. Even that, several differences can
be underlined. On one hand, in PW it was not established if the functions
$G(\mathcal{A},\mathbf{r})$ and $E_{1}(\mathcal{A},\mathbf{r})$
are valuations (also called additive functions) or not. On the other
hand, the integration domain of the integral solved in PW is not the
complete space. Furthermore, the question about the continuity or
monotonic behavior of both functions on $\mathbb{P}$ (see p. 153
in Ref. \cite{Klain1991} and pp. 211 and 253 in Ref. \cite{Schneider2008})
is open. Despite these differences, the connection of our results
with various formulae from integral geometry and convex bodies, theories
is evident. For example, some parts of the presented formulation resemble
to Steiner's formula for the measure of the Minkowski sum of a convex
polytope and a ball (see Theorem 9.2.3 in p.122 of \cite{Klain1991}),
while the solved integrals are similar to certain integrals over functions
that depend on the distance between a point and the boundary of a
convex set. (p.132 of \cite{Klain1991} and p. 258 of \cite{Schneider2008}).
These connections should deserve a deeper analysis.
\begin{acknowledgments}
This work was supported by Argentina Grants CONICET PIP-0546/10, UBACyT
20020100200156 and ANPCyT PICT-2011-1887.\bibliographystyle{apsrev}
\end{acknowledgments}

\appendix

\section{{\normalsize Series Expansions of the EOS\label{sec:AppendixB}}}

We present here the expressions for several series expansions of thermodynamic
magnitudes. In the following Eqs. the series were truncated to the
next order that were written. In the derivations we make intensive
use of relations expressed in Eqs (\ref{eq:funcset01}) to (\ref{eq:funcset05}).
From Eqs. (\ref{eq:GamInf}) and (\ref{eq:Adsorp}) we found $\gamma_{\infty}(\rho_{b})$
\begin{equation}
\beta\gamma_{\infty}(\rho_{b})=a_{2}\rho_{b}^{2}+(a_{3}-4a_{2}b_{2})\rho_{b}^{3}+\left(a_{4}-6a_{3}b_{2}+20a_{2}b_{2}^{2}-6a_{2}b_{3}\right)\rho_{b}^{4}\:.\label{eq:AppB01}
\end{equation}
Besides, $-\beta\mathcal{T}_{\infty}(\rho_{b})$ is identical to the
right hand side of Eq. (\ref{eq:AppB01}) but with the replacement
$a_{i}\rightarrow c_{i}^{\textrm{e}}$. The same applies to $-\beta\nu(\rho_{b})$
with the replacement $a_{i}\rightarrow c_{i}^{\textrm{v}}$ and also,
to $\beta\bar{\gamma}(\rho_{b})$ with $a_{i}\rightarrow a_{i}-c_{i}^{\textrm{e}}L_{\textrm{e}}A^{-1}-c_{i}^{\textrm{v}}N_{\textrm{v}}A^{-1}$.
From Eqs. (\ref{eq:GamInf}) and (\ref{eq:Adsorp}) we found the relation
between surface tension and adsorption 
\begin{equation}
\beta\gamma_{\infty}(\Gamma)=-\frac{1}{2}\Gamma+\frac{a_{3}}{4\sqrt{2}(-a_{2})^{3/2}}\Gamma^{3/2}+\frac{\left(-9a_{3}^{2}+8a_{2}a_{4}\right)\Gamma^{2}}{32a_{2}^{3}}+\frac{3\left(63a_{3}^{3}-96a_{2}a_{3}a_{4}+32a_{2}^{2}a_{5}\right)\Gamma^{5/2}}{256\sqrt{2}(-a_{2})^{9/2}},\label{eq:AppB02}
\end{equation}
which shows a non-analytic dependence between fluid-wall surface
tension and the surface adsorption in $\Gamma=0$. Besides, $-\beta\mathcal{T}_{\infty}(\Gamma_{\textrm{e}})$
is identical to the right hand side of Eq. (\ref{eq:AppB02}) but
with the replacement $a_{i}\rightarrow c_{i}^{\textrm{e}}$, while
the same applies to $-\beta\nu(\Gamma_{\textrm{v}})$ with the replacement
$a_{i}\rightarrow c_{i}^{\textrm{v}}$ and also to $\beta\bar{\gamma}(\bar{\Gamma})$
with $a_{i}\rightarrow a_{i}-c_{i}^{\textrm{e}}L_{\textrm{e}}A^{-1}-c_{i}^{\textrm{v}}N_{\textrm{v}}A^{-1}$.
Furthermore, from Eqs (\ref{eq:GamInf}) and (\ref{eq:Fluct}) we
found
\begin{equation}
\beta\gamma_{\infty}(s)=\frac{s}{2}-\frac{3}{16}\frac{a_{3}}{a_{2}^{3/2}}s^{3/2}+\frac{\left(81a_{3}^{2}-64a_{2}a_{4}\right)s^{2}}{256a_{2}^{3}}\:.\label{eq:AppB03}
\end{equation}
The same general ideas applied above in relation with Eqs. (\ref{eq:AppB01})
and (\ref{eq:AppB02}) enable to obtain e.g. $-\beta\mathcal{T}_{\infty}(s_{\textrm{e}})$
and $-\beta\nu(s_{\textrm{v}})$. Fluctuation and density in the bulk,
taken from Eqs. (\ref{eq:Fluct}) and (\ref{eq:Adsorp}), relates
by
\begin{equation}
\sigma_{b}^{2}=V\rho_{b}\left[1+2b_{2}\rho_{b}+\left(-8b_{2}^{2}+6b_{3}\right)\rho_{b}^{2}+\left(40b_{2}^{3}-48b_{2}b_{3}+12b_{4}\right)\rho_{b}^{3}\right]\:,\label{eq:AppB04}
\end{equation}
while to obtain $\sigma_{N}^{2}$ we must replace $\rho_{b}\rightarrow\rho$
and $b_{i}\rightarrow b_{i}-a_{i}AV^{-1}+c_{i}^{\textrm{e}}L_{\textrm{e}}V^{-1}+c_{i}^{\textrm{v}}N_{\textrm{v}}V^{-1}$,
in this case $V\rho=N$. On the other hand, $s(\rho_{b})$ is 
\begin{equation}
s(\rho_{b})=4a_{2}\rho_{b}^{2}+(9a_{3}-16a_{2}b_{2})\rho_{b}^{3}+\left(16a_{4}-54a_{3}b_{2}+80a_{2}b_{2}^{2}-24a_{2}b_{3}\right)\rho_{b}^{4}\:,\label{eq:AppB05}
\end{equation}
while, $s_{\textrm{e}}(\rho_{b})$ can be obtained through the replacement
$a_{i}\rightarrow c_{i}^{\mathbf{e}}$, while $s_{\textrm{v}}(\rho_{b})$
can be obtained through the replacement $a_{i}\rightarrow c_{i}^{\mathbf{v}}$.
To obtain $\bar{s}(\rho)$, we replace $\rho_{b}\rightarrow\rho$
and $a_{i}\rightarrow a_{i}-c_{i}^{\textrm{e}}L_{\textrm{e}}A^{-1}-c_{i}^{\textrm{v}}N_{\textrm{v}}A^{-1}$
in Eq. (\ref{eq:AppB05}).
\end{document}